\documentclass[a4paper,11pt]{article}

\usepackage{amsmath, amsthm, amssymb, amsfonts}
\usepackage{mathtools}
\usepackage{hyperref}
\usepackage{graphicx}
\usepackage{dsfont}
\usepackage{fullpage}
\usepackage{color}
\usepackage{soul} 
\usepackage[title]{appendix}
\usepackage{tikz}
\usetikzlibrary{patterns}
\usetikzlibrary{shapes.multipart}
\usetikzlibrary{arrows}
\usepackage{appendix}

\theoremstyle{definition}

\definecolor{labelkey}{cmyk}{.4,.2,0,0}

\newcommand{\be}{\begin{equation}}
\newcommand{\ee}{\end{equation}}
\newcommand{\bea}{\begin{eqnarray}}
\newcommand{\eea}{\end{eqnarray}}
\newcommand{\nn}{\nonumber}

\usepackage{titlesec}
\titleformat{\section}{\large\bf}{\thesection}{1em}{}
\titleformat{\subsection}[runin]{\bf}{\thesubsection}{1em}{}

\usepackage{authblk}

\makeatletter
\newcommand\appendix@section[1]{%
  \refstepcounter{section}%
  \orig@section*{Appendix \@Alph\c@section: #1}%
  \addcontentsline{toc}{section}{Appendix \@Alph\c@section: #1}%
}
\let\orig@section\section
\g@addto@macro\appendix{\let\section\appendix@section}
\makeatother

\title{\bf \large Equivalence of mean-field avalanches and branching diffusions: \\From 
the Brownian force model to the super-Brownian motion}

\author[1]{Pierre Le Doussal}
\affil[1]{\normalsize Laboratoire de Physique de l'\'Ecole Normale Sup\'erieure, ENS, Universit\'e PSL, CNRS, Sorbonne Universit\'e, Universit\'e de Paris, 75005 Paris, France}

\date{}

\begin{document}

\maketitle

\begin{abstract}
{\normalsize We point out that the mean-field theory of avalanches in the dynamics of elastic interfaces, the so-called
Brownian force model (BFM) developed
recently in non-equilibrium statistical physics, is equivalent to the so-called super-Brownian motion (SBM)
developed in probability theory, a continuum limit of branching processes related to {\it space-embedded} Galton-Watson trees. In particular the exact solvability property recently
(re-)discovered from the field theory in mean-field avalanches (the "instanton equation") maps onto the
so-called Dawson-Watanabe 1968 duality property. In the light of this
correspondence we compare the results obtained independently in the two fields,
and transport some of them from one field to the other. In particular, we discuss
a scaling limit of the branching Brownian motion which maps onto the continuum
field theory of mean-field avalanches.}
%
\end{abstract}


\newpage

{\pagestyle{plain}
 \tableofcontents
\cleardoublepage}

\section{Introduction}

{\,}

In this paper we point out a mathematical connection between 
the mean field theory of avalanches in non-equilibrium statistical physics, and the
so-called super-processes, e.g. the super-Brownian motion (SBM) in
probability theory, as well as their common connections to non-linear
partial differential equations (NLPDE). The connection is via the general theory of
branching processes. Although descriptions of avalanches in terms of branching processes
have been studied before, see e.g. \cite{BranchingZapperi95,DSFisher1998,LucillaReview2016,ETAS,SornetteETAS2002}, 
the precise mathematical connection pointed out here, has, to
our knowledge, not been explored. 
\\

The simplest branching process is the Bienaym\'e-Galton-Watson (BGW) process \cite{Bienayme,WG,Harris63} where the $M_n$
individuals alive at generation $n$ each independently gives rise to $k$ children,
$k \geq 0$ being a random integer drawn from a distribution $p_k$, with $\sum_{k \geq 0} p_k=1$, i.e.
$M_{n+1}=\sum_{j=1}^{M_n} k_j$ with i.i.d $k_j$. The case
$\mu:=\sum_{k \geq 0}  k \, p_k<1$ is subcritical (extinction), $\mu >1$ is supercritical (explosion)
and $\mu=1$ is critical. Feller showed \cite{Feller51} that for large populations
one can obtain a continuum limit of the (nearly critical) BGW process. 
This limit, the Feller process, is identical, at criticality, to the 
squared Bessel process with ${\sf d}=0$, denoted BESQ$^0$
(for integer dimensions ${\sf d}$, BESQ$^{\sf d}$ is the square of the ${\sf d}$-dimensional Brownian 
motion ${\bf B}_t$, i.e. the process $Y_{\sf d}(t)={\bf B}_t^2$). It is also known in financial mathematics 
as the Cox-Ingersoll-Ross (CIR) diffusion \cite{CIR,CIR-Yor,Hawkes-CIR}, the unique solution of the stochastic equation
\be  \label{eq:CIR}
dX_t=({\sf d} + \gamma X_t) dt + 2 \sqrt{|X_t|} dB_t
\ee
where $B_t$ is the standard (one dimensional) Brownian motion. This process is 
related to the BESQ$^{\sf d}$ process as $X_t=e^{\gamma t} Y_{\sf d}(\frac{1}{\gamma}(1-e^{-\gamma t}))$, identical to it for $\gamma=0$. Here 
$\gamma$ measures the
distance to criticality. More details are recalled in Appendix \ref{app:GW}.
\\

It turns out that the same continuum model emerged in physics, completely independently,
from experimental studies of the Barkhausen noise in ferromagnets, 
which is caused by the avalanche motion of domain walls.
There $X_t$ represents the instantaneous velocity of the center of mass 
of the domain wall. The model \eqref{eq:CIR} was proposed by 
Alessandro, Beatrice, Berlotti and Montorsi (ABBM) \cite{AlessandroBeatriceBertottiMontorsi1990,AlessandroBeatriceBertottiMontorsi1990b,Colaiori2008}
as a phenomenological toy model (i.e.
for a single degree of freedom). In that context, ${\sf d}$ is the driving velocity
and $\gamma=-m^2 \leq 0$ the effective mass, where $- m^2 X_t$ the restoring force. Recently, we went beyond the
center of mass description, to include the internal space dependence
for the domain wall, modeled as a driven interface in a random medium. We
derived \cite{LeDoussalWiese2011a,LeDoussalWiese2012a}, from the (first principle) field theory description of this problem,
also called functional RG (FRG) \cite{LeDoussalWiese2008c,LeDoussalWiese2011b},
a mean-field theory for avalanches, which includes the spatial structure
\cite{ThieryLeDoussalWiese2015,Delorme2016,DobrinevskiPhD,ThieryPhD,DobrinevskiLeDoussalWiese2014a,ThieryShape,PLDLong2021}. This theory is called the Brownian force model (BFM). 
\\

On the other hand, (discrete) BGW and (continuum) 
Feller processes describe particle splitting but do not involve space.
One well known extension of the BGW process including space, is the branching
random walk and its semi-continuum limit, the branching
Brownian motion (BBM), where particles independently perform random walks (respectively Brownian motion)
die and branch \cite{Moyal1962,sawyer79,McKean75,Bramson78}. Similarly, extensions of the Feller process including space
are called {\it superprocesses} (SP) \cite{Etheridge}, a prominent one being the 
Dawson-Watanabe SP (DWSP) \cite{Watanabe68}, also called super Brownian motion (SBM)
\cite{Perkins1,SladeSBMReview,PerkinsDWSPReview2002}.
It can be constructed as the weak limit of the rescaled BBM.
Superprocesses require the concept of measure-valued
Markov process, and were originally constructed (starting with Watanabe)
as branching processes in abstract spaces. Another example is the Fleming-Viot SP \cite{Etheridge,PerkinsFlemingViot2019},
where the total population is constant (it is in a sense the "angular part" of DWSP).
In general, the space is $\mathbb{R}^d$, or more abstract, such as the genetic space. 
Finally, the historical process decomposes these measures according to ancestral history.
\\

It is thus natural to ask about possible connections between 
the mean-field theory of avalanches (and beyond), and space-embedded
BGW trees, or superprocesses, and to explore their consequences
for both domains. This is the topic of this paper. We will recall each
side, then discuss the links, and finally mention some consequences.
In several respects the present paper has the character of a review, as we will 
quote results in both domains and compare them. This hopefully brings new
information to each of these fields, which until now have developed quite separately. 
There will be no attempt at mathematical rigor, for which
we refer to the quoted literature. 
\\

The outline is as follows. We first recall in Sec. \ref{sec:BFM} 
the mean-field theory of avalanches, i.e. the Brownian force model. In Sec. 
\ref{sec:SBM} we recall what is the super-Brownian motion. In Sec. 
\ref{sec:relations} we detail the relations between the two (BFM and SBM).
We recall in Sec. \ref{sec:BBMtoSBM} how one goes from the branching Brownian
motion (BBM) to the SBM. In Sec. \ref{sec:observables} we
recall the main observables of interest in each field, and some of the results
(some being very recent), and compare them.
Finally, in Sec. \ref{sec:extension} we briefly review the superprocesses
(which generalizes the SBM) and the related topic of avalanches for interfaces in 
presence of long range elasticity. In the Appendices we provide further information
on branching processes and avalanches, review and compare some further results in both fields,
and give some details of the derivations. 

\section{Mean-field theory of avalanches: the Brownian force model}
\label{sec:BFM} 

Consider an elastic interface defined by a scalar height field $u(x,t)$, 
with internal coordinate $x \in \mathbb{R}^d$,
evolving in a random medium according to the equation of motion
\be
\eta \partial_t u(x,t)= (\nabla_x^2 - m^2) u(x,t) + F\!\left(u(x,t),x\right) +  f(x,t) 
\label{BFMpos1}
\ee
with friction coefficient $\eta$. When $m>0$ the driving force is written $f(x,t)=m^2 w(x,t)$, 
corresponding to driving by a quadratic potential of curvature $m$ centered at $w(x,t)$.
In \eqref{BFMpos1} the elastic forces are short-ranged (modeled by the $d$-dimensional Laplacian $\nabla_x^2$) but the theory extends to long range elasticity.
Here $F(u,x)$ is a quenched random force field representing the disordered medium.
It is shown \cite{Middleton1992,MacKay,RossoPhD} that under monotonous forward driving $\dot f(x,t) \geq 0$
the interface moves only forward \cite{footnote2} $\dot u(x,t) \geq 0$
(we use interchangeably $\partial_t u$ or $\dot{u}$ for time derivatives), and 
undergoes avalanche motion, that is jerky, intermittent motion. 
For the most realistic applications $F(u,x)$ has short range correlations, a difficult problem, which can be controled using the functional RG (FRG) approach of disordered elastic systems
\cite{DSFisher1986,NattermannStepanowTangLeschhorn1992,NarayanDSFisher1992b,NarayanDSFisher1993a,ChauveLeDoussalWiese2000a,LeDoussalWieseChauve2002,LeDoussalWieseChauve2003,LeDoussal2006b,LeDoussal2008,ThieryStaticCorrelations}.
However it was shown, using the FRG \cite{LeDoussalWiese2011a, LeDoussalWiese2012a,LeDoussalWiese2008c,LeDoussalWiese2011b},
that the mean-field theory of avalanches
corresponds to choosing the random forces $F(u,x)$ as a collection of independent one-sided Brownian motions in the $u$ direction, the so-called Brownian force model (BFM) \cite{LeDoussalWiese2011b,LeDoussalWiese2012a}, i.e. a centered Gaussian field with correlator
\be
\overline{F(u,x)F(u',x')}=2 \sigma \delta^d(x-x')\min(u,u')\ . \label{defF} 
\ee
where here the overbar denotes the expectation value.
Two important properties of the BFM model for monotonous driving were obtained in \cite{DobrinevskiLeDoussalWiese2011b,LeDoussalWiese2012a,ThieryLeDoussalWiese2015}.
First, Eqs. \eqref{BFMpos1}-\eqref{defF} imply, 
from the Markov property of the Brownian
motion, that
the velocity field $\dot{u}(x,t)$ obeys the following 
stochastic differential equation (SDE) (in the 
Ito sense)\,:
\be
\eta \partial_t \dot{u}(x,t)= (\nabla_x^2-m^2) \dot{u}(x,t) 
+ \sqrt{2 \sigma \dot{u}(x,t)} \,\xi(x,t) + \dot f(x,t) 
\label{BFMdef1}
\ee
where $\overline{\xi(x,t) \xi(x',t')}=\delta^d(x-x') \delta(t-t')$ is a space time white noise.
Second, for an initial condition $\dot u(x,t=0^-)=0$ and any forward
driving $\dot f(x,t) \geq 0$, any average $G[\lambda]$ of an exponential of a linear
function of the velocity field is obtained as
\be
G[\lambda] : = \overline{\exp \left( \int d^d x \int_0^{+\infty} dt  \, \lambda(x,t) \dot u(x,t) \right) } = 
\exp \left( \int d^d x \int_0^{+\infty} dt  \, \dot f(x,t) \tilde{u}^\lambda(x,t) \right) 
\label{generating}
\ee
where $\lambda(x,t)$ is any function such that the l.h.s. side exists.
Here the expectation value is for solutions of \eqref{BFMdef1} over $\xi$, but extends to 
solutions of \eqref{BFMpos1} over $F$ under certain conditions of preparation
of the initial state $u(x,t=0)$, see e.g. \cite{DobrinevskiLeDoussalWiese2011b} and Appendix \ref{app:FRG}. 
The function $\tilde u^\lambda(x,t)$ is the solution of the "instanton equation"
\cite{LeDoussalWiese2011a,LeDoussalWiese2012a}
\be
(\eta \partial_t + \nabla_x^2 - m^2 ) \tilde{u}(x,t) + \sigma \tilde{u}(x,t)^2 = - \lambda(x,t) 
\label{instanton}
\ee
continuously connected to $\tilde u^{\lambda=0}=0$, and some vanishing conditions
at infinity \cite{footnoteC}.
The property \eqref{generating} relates mean-field avalanches to
solving a NLPDE, and has allowed to calculate exactly a number of avalanche observables
for the BFM (some will be recalled below). The BFM is also the starting point for a field theoretic dimensional expansion beyond
mean field (i.e. for short range disorder) in powers of $\epsilon = d_c-d$. More details are
given in Appendix \ref{app:FRG}. 
\\

Remarkably, it turns out that Eq. \eqref{BFMdef1} and some properties related to \eqref{generating}-\eqref{instanton}, obtained here via field theory methods,
appeared a long time ago (1968!) in an a-priori quite different field
and topic, the study of continuum limits of branching processes in probability
theory, specifically the SBM. 

\section{The super Brownian motion}
\label{sec:SBM} 

The SBM 
is a continuous measure-valued Markov process, noted $\rho_t$ (usually noted $X_t$).
$\rho_t(A)$ denotes the total weight of set $A$. In some cases (see below) it can be written as
$\rho_t(dx) = \rho(x,t) dx$. The SBM models the evolution over time $t$ of a distribution of mass. 
It is the universal limit of rescaled branching random walks, or space-embedded BGW trees. 
One example is as follows \cite{Perkins1}.
Start $O(N)$ particles in $\mathbb{R}^d$, all evolving independently.
With rate $\tilde a N$ a particle at $x$ dies, with rate $\tilde b N= \tilde a N + \gamma$ the particle at $x$ gives birth to a particle at $x+W/\sqrt{N}$, where $W$ is centered Gaussian of covariance 
$\overline{W_i W_j}= 2 D \delta_{ij}/\tilde b$. 
Consider the "empirical measure", $\rho_t^N$, 
such that $N \rho_t^N(A)$ is the number of particles in a set $A$, i.e. loosely
$\rho^N(x,t) = \frac{1}{N} \sum_{i=1}^{M(t)}  \delta(x-x_i(t))$ where $M(t)$ is the number of particles
alive at time $t$ and $x_i(t)$ their positions \cite{footnoteN}. 
The scaling 
is such that the lineage of a particle alive at time $t$ is Brownian motion in the limit $N \to \infty$. 
The mathematical statement \cite{Watanabe68,Perkins1} is that if, as $N \to \infty$, $\rho_0^N$ converges to a measure $\rho_0$  
then $\rho_t^N$ converges to a time-dependent measure $\rho_t$, which is a continuous measured-valued
Markov process, depending only on $(\rho_0, \tilde b, \gamma, D)$. This process is called the SBM. There are many other
discrete versions (with combinations of diffusion and branching) which converge to the same universal limit as long as
they are near-critical. The SBM is a particular case (the simplest) of more general 
{\it super-processes}, and is called the Dawson-Watanabe super-process, see e.g. \cite{Etheridge}.
\\

Two important results in probability theory are as follows. First \cite{Reimers89,Konno-Shiga88}
in space dimension $d=1$ the SBM admits a density
(with respect to the Lebegue measure), i.e. $\rho_t(dx) =\rho(x,t) dx$, which is the
unique solution of the SPDE
\be
\partial_t \rho(x,t) = D \nabla_x^2 \rho(x,t) + \gamma \rho(x,t) + \sqrt{2 \tilde b \rho(x,t)} \xi(x,t)
\label{spde1} 
\ee
where $\xi(x,t)$ is unit space-time white noise. 
The general meaning of this SPDE is by integration versus test functions,
which allows to still interpret \eqref{spde1} in $d>1$, where it is known
that there is no density (an infinite integrand on an infinitesimal set), see e.g. \cite{Etheridge}.
\\

The second remarkable result \cite{Watanabe68} is that the SBM satisfies 
\bea \label{duality1} 
\mathbb{E}_{\rho_0}\bigg[ \exp( - \langle \rho_t , \phi \rangle ) \bigg] = \exp( - \langle \rho_0 , v_t \rangle ) 
\eea 
where $\langle \mu , \phi \rangle=\int \mu(dx) \phi(x)$, $\mathbb{E}_{\rho_0}$ denotes the
expectation over all SBM paths starting at the measure $\rho_0$, and $v_t(x)=v(x,t)$
is the solution of
\be
 \partial_t v = D \nabla_x^2 v - \tilde b v^2 + \gamma v 
\quad, \quad v(x,t=0)=\phi(x) \geq 0  \label{instantonSBM}
\ee
This connection between the SBM and NLPDE's was much developed
and has allowed to derive a number of properties of the SBM \cite{Iscoe86,Iscoe88,DawsonPath1989,legallBook,LegallCourse,Legallpackingdimension,LegallOccupation,PerkinsSaintFlour}. 
The SBM is the universal limit of many models above their upper critical
dimension, for reviews see \cite{Perkins1,SladeSBMReview}. It is the limit for instance of the rescaled voter model
for $d \geq 2$ \cite{CoxVoter2000} (and its dual, the coalescing random walk), 
of rescaled critical contact processes for $d \geq 2$
\cite{DurrettContact1999} and $d \geq 4$ for the unrescaled version,
of critical oriented percolation in $d \geq 4+1$
\cite{SladePerco2003}, lattice trees for $d > 8$ 
\cite{AldousTree1993,DerbezTree1998,SladeReview1999} 
(convergence to the variant of SBM known as integrated super-Brownian excursion (ISE))
and standard percolation
(infinite incipient cluster) for $d>6$, and others \cite{CoxVolterra2003}.

\section{Relations between the BFM avalanches and the SBM}
\label{sec:relations} 

The SBM stochastic equation, Eq. \eqref{spde1} (setting $D=1$, $\tilde b=\sigma$, $\gamma=-m^2$),
identifies exactly with the BFM equation of motion, \eqref{BFMdef1} (with $\eta=1$),
where the velocity field identifies with the SBM density, i.e. $\dot u(x,t) = \rho(x,t)$. 
Most interestingly, the property \eqref{duality1} together with \eqref{instantonSBM},
sometimes called duality, see Appendix \ref{app:duality}, can be seen as a particular case of 
the identity \eqref{generating} for the BFM, together with the instanton 
equation \eqref{instanton} (denoting there $t \to \tau$), with the following
choices for the source and the driving 
\be \label{choice}
\lambda(x,\tau)= - \phi(x) \delta(\tau-t) \quad , \quad \dot f(x,\tau) = \rho(x,0) \delta(\tau) 
\ee
in which case one can check that \eqref{instanton} has for solution $\tilde u(x,\tau) = - v(x,t-\tau)$
where $v(x,t)$ satisfies \eqref{instantonSBM}. Indeed the source \eqref{choice} 
implies $\tilde u(x,t=\tau) = - \phi(x)$, since $\tilde u(x,\tau)=0$ for $\tau>t$ (from
the boundary condition $\tilde u(x,+\infty)=0$). The type of driving 
$\dot f(x,\tau) = f(x) \delta(\tau)$ corresponds to submit the interface to a kick in force at time zero, and in the relation
\eqref{choice} its amplitude $f(x)$ is precisely the initial SBM density. More details are given in Appendix 
\ref{app:relation}. 
\\

At this stage however the BFM result appears more general, since it allows to
study any time dependent source and (positive) driving.
However one can find in 
\cite{Iscoe86} an analogous result for the SBM 
(for so-called weighted occupation times) 
corresponding to an arbitrary source $\lambda(x,t)$, but still with
only a kick driving. In the BFM one can study more general 
time dependent driving $\dot f(x,t)$ (e.g. as in \cite{DobrinevskiLeDoussalWiese2011b}).
It turns out that this corresponds
to what is known as {\it superprocesses with immigration.}
In the BGW process it simply means that the number $M_n$ of individuals alive at 
(discrete) time $n$ now obey the recursion 
\be 
M_{n+1}=\sum_{j=1}^{M_n} k_j + m_n
\ee
where $m_n$ new individuals enter at generation $n$, see e.g. \cite{Pakes1972}. It is 
clear from \eqref{BFMdef1} that in the continuum setting
it translates into a local time-dependent source 
$d \rho(x,t) = \dot f(x,t) dt$ (while it may take a more abstract 
form in the probability literature, see e.g. \cite{immigration,LiImmigration}).

\section{From the branching Brownian motion to the SBM and to the BFM}
\label{sec:BBMtoSBM} 

Consider the branching Brownian motion (BBM) in space dimension $d$, where independent 
particles undergo diffusion with coefficient $D$ ($D=1/2$ for standard Brownian motions) and
die at rate $V$ by producing
$k \geq 0$ offsprings where the function 
\be \label{mechanism} 
\Phi(z)=\sum_{k \geq 0} p_k z^k \quad , \quad \Phi(1)=1
\ee 
is often called the branching mechanism. Then it
is well known \cite{McKean75,Etheridge} (and recalled in Appendix \ref{app:BBMcalc}) 
how to calculate expectation values of any product of a test function $0 \leq h(x) \leq 1$
over particle positions $x_j(t)$,
$j=1,.. M(t)$ alive at time $t$, with initial condition a single particle at $x$ at $t=0$. One has that
\cite{footnote0}
\be
g(x,t)= \mathbb{\mathbb{E}}_{\{x\}} \left[ \prod_{j=1}^{M(t)} h(x_j(t)) \right]  \label{defg} 
\ee 
is the unique solution for $t \geq 0$ of 
\be  \label{kpp1} 
 \partial_t g = D \nabla_x^2 g + V( \Phi(g)-g ) \quad, \quad g(x,t=0)=h(x) 
\ee
For the branching in two, $\Phi(z)=z^2$ and
this is the celebrated FKPP equation \cite{Fisher37,KPP,McKean75,Bramson78}.
A standard choice to study the vicinity of the critical case is the binary branching 
\be  \label{binary} 
V (\Phi(z)-z) = b z^2 + a - (b+a) z
\ee 
where $b$ is the branching rate and $a$ the death rate, criticality being realized when $b=a$. 
\\

Suppose now that there are $M(0)$ initial particles, at
positions $x_j(0)$. Since they evolve independently, the analog expectation value in the r.h.s. of 
\eqref{defg} is now equal to $\prod_{j=1}^{M(0)} g(x_j(0),t)$, where $g$ satisfies
also Eq. \eqref{kpp1},
which can be rewritten in a form similar looking to \eqref{duality1} 
\be \label{gener2} 
\exp( - N \int d^dx \rho(x,0) \ln g(x,t) ) 
= \mathbb{E}_{\{x_j(0)\}}\bigg[ \exp( N \int d^dy \rho(y,t) \ln h(x)) \bigg] 
\ee
where $N \rho(y,t)=\sum_{j=1}^{M(t)} \delta(y-x_j(t))$, $N$ here being a parameter \cite{footnoteN}. 
Now the SBM is obtained in the scaling limit of a large number of particles, $M(0)=O(N)\gg 1$, with
large branching and death rates which barely compensate each
others \cite{Etheridge}. More precisely one chooses $V = N V'$ and the test function as
\be
h(x) = 1 - \frac{1}{N} \phi(x)  
\ee
In the limit $N \gg 1$, the solution of \eqref{kpp1} takes the form
\be \label{gN} 
g(x,t)= 1 - \frac{1}{N} v(x,t) + O(\frac{1}{N^2})
\ee
Inserting \eqref{gN} in \eqref{kpp1}, using that $\Phi(1)=1$ by definition
and tuning near criticality $\Phi'(1)=1 + \frac{\gamma}{V' N} + O(\frac{1}{N^2})$ 
we see that $v(x,t)$ satisfies, for $N \to +\infty$, exactly the
equation \eqref{instantonSBM}, with $\tilde b=\frac{1}{2} V' \Phi''(1)=b/N$.
Taking the large $N$ limit of \eqref{gener2} we obtain exactly
\eqref{duality1}. Hence $\rho_t(dx)=\rho(x,t) dx$ follows the SBM evolution. 
Thus the SBM measure $\rho_t$ provides a continuous description of the empirical density
of a large number of branching Brownian particles with high branching and death rates. 
\\

One can ask whether one can connect directly the BBM to the avalanches, i.e. to the BFM.
Since the BFM instanton equation \eqref{instanton} contains more general information, one can
ask whether an analog formula exist for the BBM, over weighted occupation
time. Indeed it is the case, and one shows, see Appendix \ref{app:multitime}, that for a source function $\Lambda(x,t) <0$ 
\be \label{instBBM1} 
\mathbb{E}_{\{x_j(0)\}}\left[ \exp\left( \int_0^{+\infty} dt \sum_{j=1}^{M(t)} \Lambda(x_j(t),t) \right) \right]
= \exp\left(  \sum_{j=1}^{M(0)} \ln \tilde g(x_j(0),0) \right)
\ee
where $\tilde g(x,\tau)$ obeys the equation
\bea \label{instBBM2} 
- \partial_\tau \tilde g = D \nabla_x^2 \tilde g + V ( \Phi(\tilde g) - 
\tilde g) +  \Lambda \tilde g 
\eea
which one solves backwards in time starting from $\tilde g(x,\tau=+\infty)=1$ down to
$\tau=0$. Inserting $\Lambda(x,\tau)=\log h(x) \delta(\tau-t)$ one recovers
Eq. \eqref{defg}, i.e. the single time average. The formulas \eqref{instBBM1},\eqref{instBBM2}
also allow to compute multi-time observables for the BBM, as detailed in
Appendix~\ref{app:multitime}. 
\\

The limit from the BBM to the BFM is obtained for $M(0)=O(N)\gg1$ as follows. 
The empirical density of particles become the velocity field (also called activity field)
\be 
\rho(x,t)=\frac{1}{N} \sum_j \delta(x-x_j(t)) \to \dot u(x,t)
\ee 
Setting $\tau = t$, $\Lambda(x,t)=\frac{1}{N} \lambda(x,t)$,
and $\tilde g(x,t) = 1 + \frac{1}{N} \tilde u(x,t) + O(1/N^2)$, the BBM equations
\eqref{instBBM1},\eqref{instBBM2} recover, in the limit of large $N$, 
the BFM equations \eqref{generating},\eqref{instanton}, with $\eta=1$
and a kick driving $\dot f(x,t)=\dot u(x,0) \delta(t)$. To obtain the BFM with a 
general driving $\dot f(x,t)$, one considers the BBM with immigration, i.e.
adding an atomic measure $N d \rho(x,t)=\dot f(x,t) dt$ in each time slice $dt$.
The BBM equation \eqref{instBBM1} then generalises, in presence of immigration,
to 
\be
\mathbb{E} \left[ \exp(\left( \int_0^{+\infty} dt \sum_{j=1}^{M(t)} \Lambda(x_j(t),t) \right) \right]  
= \exp \left( \int_0^{+\infty} dt d^dx \dot f(x,t) \ln \tilde g(x,t) \right) 
\ee

\medskip

\section{Observables and results} 
\label{sec:observables} 

We now discuss the main observables in each system (BFM on one hand and SBM or BBM on the other hand)
and compare some known results.

The equation of motion \eqref{BFMpos1} leads to {\it avalanches}, i.e. upon a
kick $\dot f(x,t) = f(x) \delta(t)$ the interface prepared initially at rest, takes initial velocity $\dot u(x,0^+)=f(x)$, 
then moves, and eventually stops for
$m^2 \geq 0$, the case of main interest here. 
In the BFM model 
\eqref{defF} any small kick leads to an avalanche (most of them being small) of finite
duration $T$. One defines the local and total size of the avalanche 
\be
S(x)=  \int_0^{+\infty} dt \, \dot u(x,t) \quad , \quad S = \int d^d x \, S(x)
\ee
and the total instantaneous velocity $\dot u(t)=\int d^d x  \, \dot u(x,t)$. 
Clearly the avalanches in the BFM map to a subcritical ($m^2>0$)
or a critical ($m^2=0$) SBM. The BFM with the kick driving $\dot f(x,t)=f(x) \delta(t)$
maps to the SBM with an initial measure at time $t=0^+$ given by $\rho(x,0) dx=f(x) dx$.
Since $\dot u(x,t)$ corresponds to $\rho(x,t)$, we see that $\dot u(t)$ is the total mass
$\rho_t(\mathbb{R})$ of the SBM at time $t$, i.e. the large $N$ limit of $M(t)/N$
for the BBM. The avalanche duration $T$ is the extinction time of the branching process,
always finite for $m^2 \leq 0$. Finally, the local avalanche size $S(x)$ 
corresponds to the density of the time integrated local time measure $L_x$ (see below) and 
the total avalanche size $S$ is the time integrated total mass of the SBM.  For the BBM, the total
avalanche size $S$
corresponds to the large $N$ limit of $\int_0^{+\infty} dt M(t)/N$ (in the BGW model, which is discrete in time, it would
be the total number of individuals who have lived). This is summarized in Table I. 

\begin{table}[h!]
\begin{center}
\begin{tabular}{|c|c|c|}
\hline
Avalanches (BFM) & SBM  & BBM (scaled)  \\
\hline
local inst. velocity & local mass density & empirical density \\
$\dot u(x,t)$ & $\rho(x,t)$ & $\frac{1}{N} \sum_{j=1}^{M(t)} \delta(x-x_j(t))$ \\
\hline
total inst. velocity & total mass & total particle number \\
$\dot u(t)$ & $\rho_t(\mathbb{R}^d)$ & $\frac{1}{N} M(t)$ \\
\hline
avalanche duration & extinction time & extinction time \\
\hline 
local size & local time 
& time integrated density \\
$S(x)$ & $L_x = \int_0^{+\infty} dt \rho(x,t) $ &  \\ 
\hline 
total size &  ISE total mass &  \\
$S$ & $L= \int_0^{+\infty} dt \rho_t(\mathbb{R}^d)$   & $\frac{1}{N} \int_0^{+\infty} dt M(t)$ \\
\hline
kick driving &  initial measure &  \\
$f(x,t)=f(x)\delta(t)$ & $\rho_0(dx)=f(x)dx$   & $\frac{1}{N} \sum_{j=1}^{M(0)} \delta(x-x_j(0))$ \\
\hline

\end{tabular}
\end{center}
\label{Table0} \caption{Correspondence between observables: avalanches in the Brownian force model (BFM), versus super Brownian motion (SBM) and the scaled
version of the branching Brownian motion (BBM).
}
\end{table}

An important property of the BFM is that the integral of $\dot u(x,t)$ along $d'$ of its coordinates
(that is along a subspace in $x$ of dimension $d'$) obeys the BFM equation \eqref{BFMdef1} 
in dimension $d-d'$. For $d'=d$ it means that the total velocity $\dot u(t)$ follows the ABBM model
(which can be considered as the BFM in dimension $d=0$). This property is also well known
for the SBM, and is called the additivity property: if $\rho_1$ and $\rho_2$ are two SBM's
then $\rho_1+\rho_2$ is also a SBM \cite{Perkins1}. 
\\

In the (subcritical) BFM the natural scales for the avalanche size $S$ and duration $T$ 
are $S_m=\sigma/m^4$ and $\tau_m=\eta/m^2$,
which are also the cutoffs for the avalanches at large scale. 
By rescaling space and time with powers of $m$, 
one can use dimensionless units where $S_m=\tau_m=1$. In these
units the probability density functions (PDF) for the total size $S$ and the total duration $T$ read
(see \eqref{PfS} for the same result in original units)
\be 
P_f(S)=\frac{f}{2\sqrt{\pi} S^{3/2}} e^{- (S- f)^2/(4 S)} \quad , \quad 
P_f(T<t)= e^{-f/(e^t-1)}
\ee
where here and below we denote $f=\int d^d x f(x)$ the total
amplitude of the kick. Here $f>0$ acts as a small scale cutoff, since 
avalanches smaller that $S < f^2$ are strongly suppressed. 
These two distributions show power law decay $S^{-3/2}$ and $T^{-2}$ at intermediate scales,
and at all scales in the limit $f \to 0$ and in the critical massless limit $m \to 0$, see
\eqref{PfS}. The joint 
PDF's $P_f(S,T)$, $P_f(\dot u(t),T)$, $P_f(\dot u(t),S)$ and $P_f(\dot u(t_1),\dot u(t_2))$
were calculated in \cite{DobrinevskiLeDoussalWiese2011b} upon inverse Laplace transformation 
of \eqref{generating}, solving
\eqref{instanton} with space independent sources,
respectively $\lambda(x,t)=\lambda$ 
(for $S$), $\lambda(x,t)=\lambda \delta(t-t_1)$ (for $\dot u(t_1)$ or for $T$)
and combination thereof. All these results can be immediately translated into results for
the corresponding SBM observables, $f=\dot u(0^+)$ being the initial total mass.
Note that the full joint PDF of the field $\{ S(x) \}_{x \in \mathbb{R}^d}$ was obtained in 
\cite{ThieryLeDoussalWiese2015}. We have reproduced its formula in \eqref{PSX} 
in Appendix \ref{app:observables}. We are not aware of a similar result 
for the SBM. 
\\

In the special case of a time-independent driving at a fixed velocity, $\int d^dx \dot f(x,t) =v$, 
it is known that there is a critical velocity $v_c$ such that the 
total instantaneous velocity $\dot u(t)$ of the interface vanishes an infinite
number of times for $v<v_c$, and only a finite number of time for $v>v_c$
(in which case there is an avalanche of infinite duration) see Appendices \ref{app:GW}
and 
\ref{app:immigration}. 
Hence, in the SBM (and the BBM) increasing the immigration rate
leads to similar transitions (i.e. recurrent versus transient extinctions),
\\

For avalanches it is natural to define {\it densities} as the response to an infinitesimal kick.
For instance the avalanche total size density is 
$\partial_f P_f(S)|_{f=0} = \frac{1}{2\sqrt{\pi} S^{3/2}} e^{- S/4}$, but densities can
be defined for any observable. 
They are unnormalized measures containing information for events with $S=O(1)$ while 
most avalanches are vanishingly small $S=O(f^2) \to 0$. These densities
are expected to be universal, e.g. driving the interface with uniform velocity $f(x,t)/m^2 =v t$ 
generates the same densities in the limit  of so-called quasi-static driving $v=0^+$. 
One interesting case discussed below if the {\it local kick driving} 
\be \label{xs} 
\dot f(x,t) = f(x) \delta(t) \quad , \quad f(x) = f \delta(x-x_s)
\ee 
which forces the "seed" of the avalanche to be at $x=x_s$. In the SBM the local kick driving
amounts to consider an initial measure localized at $x=x_s$
(i.e. the scaling limit of a tree started from $x=x_s$). Considering 
an infinitesimal $f$ allows to define the densities of observables for avalanches conditioned
to their seed being at $x_s$. 
It is thus natural to surmise that avalanche densities associated to an infinitesimal 
local kick driving map onto the {\it canonical measure} for the SBM, 
defined, see e.g. \cite{Perkins1}, as $N_0[ \rho] = \lim_{N \to +\infty} N P[\rho_N | \rho_0 = N^{-1} \delta_{x,x_s} ]$,
which also gets rid of all the infinitesimally small clusters.

\bigskip

The spatial information about $S(x)$, and its support $\Omega=\{ x \in \mathbb{R}^d ~ \text{s.t.} ~ S(x)>0\}$, i.e. the set of points which have moved during the avalanche, is obtained by solving the instanton equation choosing static sources $\lambda(x,t)=\lambda(x)$ in Eq. \eqref{generating}-\eqref{instanton}. Consider an avalanche with seed at $x=x_s$,
i.e after the local kick driving \eqref{xs}.  To calculate the probability that a set $\Gamma$ intersects $\Omega$, one chooses $\lambda(x)=-\infty$ for $x \in \Gamma$ and $\lambda(x)=0$ for $x \in \mathbb{R}^d - \Gamma$ leading to
\be \label{prob0} 
{\rm Prob}(S(y)=0 ~ , ~ \forall y \in \Gamma) = \exp\left( f \tilde u(x_s) \right) 
\ee
where $\tilde u(x) \leq 0$ solves (here and below we focus on the critical case $m^2=0$, unless explicitly mentioned,
setting here $\sigma=1$) 
\be \label{inst2n} 
\nabla_x^2 \tilde u(x)  + \tilde u(x)^2 = 0 \quad , \quad \tilde u(x)|_{x \in \Gamma}=- \infty 
\ee
and vanishes at infinity. In space dimension $d=1$ one shows that the support of an avalanche started at $x=0$ is an 
interval $\Omega=[-\ell_1,\ell_2]$. Solving \eqref{inst2n} for $\Gamma=\{-l_1,l_2\}$ one obtains the joint CDF
\cite{Delorme2016,PLDLong2021} 
\be \label{Pl1l2n}
{\rm Prob}(\ell_1<l_1,\ell_2<l_2)= \exp\left( - \frac{6 f}{l^2} {\cal P}_0(p) \right)  ~,~ p=\frac{l_2}{l} ~,~
l=l_1+l_2
\ee
where ${\cal P}_0(z)={\cal P}(z;g_2=0,g_3=\frac{\Gamma(1/3)^{18}}{(2 \pi)^6})$
is the elliptic Weirstrass function. From \eqref{Pl1l2n} one obtains $P(\ell)$, the PDF of the total extension (or span)
$\ell = \ell_1+\ell_2$. Its density is given by 
\be 
\rho(\ell)=\frac{8 \pi \sqrt{3}}{\ell^3}
\ee
This exponent $-3$ shows up in a simpler observable, the probability that an avalanche starting at $0$ never reaches $x=y$. It is obtained solving \eqref{inst2n} 
for $\Gamma=\{y\}$, which leads to $\tilde u(x)=-6/(x-y)^2$, and from \eqref{prob0}, to 
$p(\ell_2<y)=\exp(- 6 f/y^2)$. Finally note that for a more general kick the 
r.h.s of \eqref{prob0} is replaced by 
$\exp\left(\int d^d x f(x) \tilde u(x) \right)$. 
\\

Some of these results about the spatial extent of an avalanche extend for $d>1$. 
Consider an avalanche started at $x=x_s=0$.
The probability that a point $y$ of the interface has not moved is simply (for $d<4$)
${\rm Prob}(y \notin \Omega) = e^{- f \frac{2 (4-d)}{y^2}}$, since 
$\tilde u(x)= - \frac{2 (4-d)}{|x-y|^2}$ is again solution of \eqref{inst2n}
for $\Gamma=\{y\}$. Moreover, taking $\Gamma$ to be a sphere of radius $R$,
the probability that the avalanche remains within the sphere is $e^{f \tilde u(0)/R^2}$, where 
\bea
\tilde u''(r) + \frac{d-1}{r^2} \tilde u'(r) + \tilde u^2 =0 \quad , \quad \tilde u(1)=-\infty 
\eea 
with \cite{PLDLong2021} $\tilde u(0)|_{d=1}=- \frac{3}{2} {\cal P}_0(1/2)=- 8.8475..$, 
$\tilde u(0)|_{d=2}=-12.563$, $\tilde u(0)|_{d=3}=-15.718$. Finally, 
the probability that the avalanche does not hit a cone in $d=2$ was
also computed in \cite{PLDLong2021}. 
\\

In some cases, similar results (and many more) were obtained (mostly earlier) 
in the context of SBM,
(as well as for the BBM \cite{sawyer79}) through rigorous analysis of NLPDE's.
For instance in Theor. 1 in \cite{Iscoe88} with 
$\tilde u(0)=- \frac{1}{6} (\frac{\Gamma(1/2) \Gamma(1/6)}{\Gamma(2/3)})^2 = - 8.8475..$.
It was shown in \cite[Thm 2]{Iscoe88} and \cite{LegallOccupation}
that the probability that the SBM started at $0$ ever meets a ball
of fixed radius $\epsilon$ centered at a large distance $x$ 
is $\simeq  \frac{2 (4-d)}{x^2}$
for $d<4$, $\simeq  \frac{2}{x^2 \log(x)}$ in $d=4$, 
$\simeq  \frac{c_d \epsilon^{d-4}}{x^{d-2}}$ in $d>4$, see e.g. 
\cite{footnoteHit} for some mathematical subtelties. 
\\

For the BBM taking in \eqref{instBBM1} $\Lambda(x,t)=-\infty$ for $x \in \Gamma$ and 
$\Lambda(x,t)=0$ elsewhere, Eq. \eqref{instBBM2} becomes
\bea \label{instBBM3} 
- D \partial_\tau \tilde g = \nabla_x^2 \tilde g + V ( \Phi(\tilde g) - 
\tilde g) \quad , \quad \tilde g(x,\tau)_{x \in \Gamma}=0
\eea
with $\tilde g(x,+\infty)=1$ for $x \notin \Gamma$. The probability that
no offsprings of an individual starting at $x$ will ever visit $\Gamma$ is 
$q(x)=\tilde g(x,0)$ and is thus given by the stationary solution of \eqref{instBBM3} 
\bea \label{instBBM4} 
0 = D \nabla_x^2 q + V ( \Phi(q) - q)  \quad , \quad q(x)|_{x \in \Gamma}=0
\eea
with $q(x) \to 1$ at infinity. This recovers the seminal analysis of \cite{sawyer79}. 
In space dimension $d=1$, for binary branching \eqref{binary} 
in the critical case $b=a$, and for $\Gamma=\{0\}$ one obtains for $x>0$, 
\be 
1-q(x)=p(x)=\frac{6 D}{b} (x+\sqrt{6 D/b})^{-2}
\ee
(it is called ${\cal R}(x/\sqrt{D/b})$ in \cite{SatyaBBM1}).
In the sub-critical case $\delta = a/b-1$
\bea
p(x) =  \frac{3 \delta}{2} \frac{1}{\sinh^2 \left( \frac{\sqrt{b \delta}}{2 \sqrt{D}} x + \sinh^{-1} \sqrt{\frac{3 \delta}{2}} \right)} 
\eea 
The corresponding result for the BFM \cite{PLDLong2021} (and thus for the SBM) is 
\bea
q_{\rm BFM}(x)={\rm Prob}(\ell_2 < x) =  \exp\left(  - f \frac{m^2}{\sigma} \frac{3}{2 \sinh^2(\frac{m x}{2}) } \right) 
\eea 
One can check that the two results (for the BBM and for the BFM/SBM) 
match if one considers in the BBM a large number $M \gg 1$ of initial particles at $x$, with $\delta \sim 1/M$ and
$b \sim M$. Indeed, the probability that all the particles and their descendents avoid $0$ is
$q(x)^M \simeq e^{- M p(x)} \simeq q_{\rm BFM}(x)$ 
with $M \delta = f \frac{m^2}{\sigma}$ and $m = \sqrt{b \delta/D}$.
Similarly, the distribution of the spatial extent of a BBM tree is the BBM analog of the joint PDF, $P(\ell_1,\ell_2)$, for the BFM obtained in \cite{Delorme2016}
and recalled in \eqref{Pl1l2n}. It was computed in \cite{SatyaBBM1}, leading to the same universal
tail for the PDF of the span of the BBM tree, $\sim 8 \pi \sqrt{3}/\ell^3$, as in the BFM. Observables
such as gaps between BBM particles, obtained in
\cite{SatyaBBM2,SatyaBBM3}, have however no analog in the continuum setting
of the SBM/BFM.
\\

Another interesting question is what is the fractal dimension of an avalanche? 
This question was answered in the SBM literature both for (i) the active region
at a given time $t$, which in the BFM is the set of points where the interface 
is moving with $\dot u(x,t)>0$, and in the SBM is the support of the SBM measure $\rho_t$,  
and for (ii) the total support $\Omega$ of the avalanche, 
which corresponds to the so-called "range" ${\cal R}$ of the SBM.

In the first case (i), it was shown, see e.g. \cite[Sec 6.3]{Etheridge},
that the support of the SBM measure at a given time $t$, ${\rm Supp}(\rho_t)$, 
if non-empty, has Hausdorff dimension $d_H=\min(2,d)$ (the definition of $d_H$ being recalled there).
For more detailed results see e.g. \cite{Legallpackingdimension,LegallPerkinsHausdorff1995}.
Despite that, for $d>1$ the mass of the process is located in connected components which are single points.
More precisely, defining the closed support $S(\rho_t):=\overline{{\rm Supp}(\rho_t)}$,
at any fixed $t$ and for $d \geq 2$, $S(\rho_t)$ is a.s. a Lebesgue null set \cite{PerkinsHausdorff1989}. 
For $d \geq 3$ it was shown that a typical point of $S(\rho_t)$ is
disconnected from the rest of $S(\rho_t)$, more precisely \cite[Tm1]{Tribe1991}
for a.a. $x$ in $S(\rho_t)$ its connected component is $\{x\}$.
For $d \geq 4$, $S(\rho_t)$ is totally disconnected \cite{Perkins1995}. 
Finally, a non trivial multifractal spectrum was found for the SBM
at points of low density \cite{PerkinsMultifractalSBM}. 

To study (ii) one considers the local time, or occupation time, of the SBM as a random measure,
which for $d \leq 3$ has a density $L_x^t$ \cite{Sugitani1989}. It is defined
for tests functions $\phi$, such that $\int_0^t ds \langle \rho_s ,\phi \rangle = \int d^dx \phi(x) L_x^t$. Hence the 
total local time,
$L_x: = L_x^{t=+\infty}$ is the analog of the local avalanche size, $S(x)$, as mentioned above and 
in Table I.  The {\it range} of the SBM is defined
as the closure (denoted here $\overline{\dots}$),
${\cal R} = \overline{\bigcup_{t>0} {\rm Supp}(\rho_t)}$ or as
${\cal R}=\overline{\{ x:L_x>0\}}$. Hence ${\cal R}$ is the closure of the support
$\Omega$ in the BFM. It is proved that its Hausdorff dimension is $d_H=\min(4,d)$,
see \cite{Etheridge} [Sec 6.3, Thm 6.15] and \cite{legallBook}.
\\

What is the regularity of $S(x)$ as a function of $x$ ? In space dimension $d=1$ $L_x$ is globally continuous
and it is proved in \cite[Thm 1.7]{MytnikBoundary} that the set $\{x , L_x >0 \}$ 
is an open interval $(-\ell_1,\ell_2)$ confirming the result mentioned above for the BFM. 
For space dimension $d \geq 1$ this is a delicate question and there are no results for
the BFM. For the SBM, in \cite[Thm 2.2]{MytnikBoundary} $L_x$ is proved to be 
$C^{(4-d)/2-\eta}$ Holder continuous for any $\eta>0$
for $x$ away from 0 (the seed of the avalanche) for $d \leq 3$.
For $d \leq 3$, the range of the SBM ${\cal R}$ 
has positive Lebesgue measure \cite{Hong4} and for 
$d \geq 4$ it is a Lebesgue null set with $d_H=4$ for a SBM
started as $\delta_0$ (i.e. a delta function density at $x=0$)
\cite[Thm 1.4] {DawsonPath1989}. 
 \\
 
For the BFM the mean avalanche spatial shape conditioned to a fixed total size $S$, $\langle S(x) \rangle_S$, with the seed at $x=0$, was computed in any space dimension $d$ in \cite{ThieryShape}.
In space dimension $d=1$, the same spatial shape, but conditioned to a fixed avalanche extension $\ell$, 
$\langle S(x) \rangle_\ell$, was computed in \cite{ZhuWiese2017,PLDLong2021}. 
It allows to investigate what happens near any of the two
edges of the avalanche. Conditioned to the support being $[-\ell_2,\ell_1]$,
the profile takes the scaling form
$S(x)= \ell^3 s(z)$ with $z=(x+\ell_1)/\ell$. Near the upper edge of the avalanche at $x=\ell_2$, it is found that
the local size vanishes with a cubic law,
$S(x) \simeq (\ell_2-x)^3 \sigma$, where the PDF of the random variable
$\sigma$ was obtained in \cite{PLDLong2021}. In dimension
$d>1$ one can pick a direction (i.e. some coordinate) and 
define the span $[-\ell_2,\ell_1]$ along that direction, for 
an avalanche with seed at $x=0$. An analogous observable is then the
mean shape $\langle S(x) \rangle_{\ell_2}$, conditioned
to the position of the upper span $\ell_2$. It was 
computed in \cite{PLDLong2021} and found to involve Bessel functions of index $7/2$.
The same functions also occur when computing the spatio-temporal shape 
$\langle \dot u(x,t) \rangle_{\ell_2}$ for the BBM in \cite{PLDLong2021}.
Interestingly, they also appeared in \cite{SatyaBBM4}, when computing the joint PDF of the position and time
at which a BBM tree reaches its farthest position in a given direction. 
Indeed, to obtain these three observables, one must linearize the instanton equation of the BFM 
around the special solution $\tilde u(x) = - 6/x^2$ (and the corresponding equation and solution for
the BBM). Finally, note that in Ref. \cite{SatyaBBM4} several results about 
the convex hull of the BBM were obtained. 
\\

Another result for the BFM in \cite{PLDLong2021} is that near any point outside of the support, i.e. such that $S(y)=0$,
the mean shape vanishes locally with a non trivial exponent $b_d$ which depends on the space
dimension $d$
\be 
\langle S(x) \rangle_{S(y)=0} \sim |x-y|^{b_d} \quad , \quad  b_1=4 \quad , \quad b_2=2 \sqrt{2}
\quad , \quad b_3=\frac{1}{2}(\sqrt{17}-1)
\ee
It was also found in \cite{PLDLong2021} that the density of $r_{\min}$, the
distance of closest approach of an avalanche to a point $y$ not in an avalanche 
behaves with a non trivial exponent as 
\be 
\sim 1/r_{\min}^{\gamma_d}  \quad , \quad \gamma_2=3-2 \sqrt{2} 
\quad , \quad \gamma_3=3- \frac{1}{2} (1+\sqrt{17})
\ee
\\

It turns out that similar non trivial exponents were obtained 
in the SBM context, in 
connection to the fractal dimension of the boundary of the range of the SBM (i.e. of the support of the avalanche). 
These studies distinguish $F=\partial \{x , L_x >0 \}$ and $\partial {\cal R} \subset F$ (topological boundary of the range 
${\cal R}=\overline{\{ x, L_x>0\}}$). These sets will differ if there are isolated zeroes of $L_x$, which will be in $F$
but not in $\partial R$: it is unknown if these isolated zeroes exist for $d=2,3$, but in $d=1$ one has $F=\partial {\cal R}=\{L,R\}$ 
\cite{Hong4}. Nevertheless it is proved in \cite{MytnikBoundary} and \cite{MytnikBoundary2018} that 
\bea
&& {\rm dim}(F)={\rm dim}(\partial {\cal R})=d+2-p(d) \\
&& p(1)=3 \quad , \quad p(2)=2 \sqrt{2} \quad , \quad p(3)=\frac{1+\sqrt{17}}{2} \nonumber
\eea
These exponents in fact appeared earlier in \cite{Werner1997}. There it is proved
for $d \leq 3$ that 
\be
N_\epsilon( \sup_{x \in {\cal R}} ||x|| \geq 1, 0 \notin {\cal R}) \sim_{\epsilon \to 0} \epsilon^{p(d)+2-d}
\ee
$N_\epsilon$ is the canonical measure starting from a point $\epsilon$ to the origin.
\\

%

There are many more results concerning the SBM and the BFM. Some of them are discussed,
and when possible compared, in Appendix \ref{app:observables}.

\section{Extensions}
\label{sec:extension} 

\subsection{Superprocesses and long range BFM}.
The SBM is an example of the more general class of superprocesses. These can be defined abstractly,
as continuous state branching processes \cite{Watanabe68} or constructed as limits of branching diffusions. 
They satisfy the branching property, that is the distribution $P_t(.,\rho_0^{(1)} + \rho_0^{(2)})$ 
of the (measure valued) process with initial value $\rho_0^{(1)} + \rho_0^{(2)}$,
is the sum of two independent copies of the
process with $\rho_0^{(1)}$ and $\rho_0^{(1)}$ respectively, i.e.
$P_t(.,\rho_0^{(1)} + \rho_0^{(2)}) = P_t(.,\rho_0^{(1)}) * P_t(.,\rho_0^{(2)})$. Since it extends to an arbitrary number 
$n$ of copies, it means that a superprocess is an infinitely divisible random measure. This is an extension
of the notion of infinitely divisible random variables, to measures. Superprocesses are then time homogeneous Markov processes which satisfies the branching property and with infinitely divisible distribution. There is an extension of the
Levy-Khintchine formula (which characterizes the possible characteristic functions of infinitely divisible distributions of random variables)
to Laplace transforms of random measures, see e.g. \cite[Thm1.28]{Etheridge}. This allows to 
classify superprocesses $\rho_t$, as they still satisfy the "duality" property \eqref{duality1}, 
but with an associated "instanton equation" which is more general than \eqref{instantonSBM},
see \cite{Etheridge,Dynkin} and Appendices \ref{app:superprocesses}, \ref{app:duality}. 

A canonical example of superprocess $\rho_t$ is the  $(\alpha,d,\beta)$ superprocess, 
$\alpha \in (0,2]$ and $\beta \in (1,2]$,
$\rho_t$ which is constructed as a special limit 
of branching diffusions in $\mathbb{R}^d$ where the spatial motion of individual particles is given by an $\alpha$-symmetric  
stable process, which has infinitesimal generator equal to the fractional Laplacian $\Delta_\alpha=- (-\Delta)^{\alpha/2}$, and where the branching mechanism \eqref{mechanism} 
\be 
\Phi(z) = z + \frac{1}{1+ \beta} (1-z)^{1+\beta}
\ee 
with $\Phi(1)=\Phi'(1)=1$, 
belongs to the domain of attraction of a stable law of index $1+ \beta$. For $0< \beta<1$ the 
variance of the number of children is infinite, while the mean remains unity. Here $\alpha=2$ is the standard Brownian motion and $(2,d,1)$ is the usual SBM (also called Dawson-Watanabe superprocess). 
\\

In the context of elastic systems the case $\alpha<2$ is of great interest since it models interfaces with long range (LR) elastic forces, which are ubiquitous in nature. It has thus been studied both within the mean field theory, i.e. 
the LR-BFM (with however finite variance $\beta=2$), and beyond mean field, either in numerical simulations, or within a $\epsilon=2 \alpha-d$ expansion using functional RG methods \cite{LeDoussalWiese2012a}. One important feature (see below) which was noted within the physics community (and studied mostly for $d=1$) is that the avalanches, while being
"compact objects" (meaning connected) for short-range elasticity (i.e. for Brownian diffusion), is made, in the case of 
LR elasticity, of many disconnected clusters, whose precise definition and statistics is still a subject of debate. 

\subsection{Results and observables}. The "instanton equation", analogous to \eqref{instanton}, 
associated to the superprocess $(\alpha,d,\beta)$ reads 
\be \label{instLR} 
\partial_t \tilde u + \Delta_\alpha \tilde u - m^2  \tilde u + \sigma (- \tilde u)^{1+\beta} = - \lambda(x,t) 
\ee 
Here we use alternatively the language and notations of the SBM or of the 
LR-BFM for which this equation was also studied (in the case $\beta=1$). 
We restrict below to the critical case $m^2=0$. 
\\

The equation \eqref{instLR} was studied in \cite[Sec 5]{Iscoe88} in the case
of Brownian diffusion $\alpha=2$, and for branching mechanism $\beta \in (0,1)$. 
The natural extension of the special static solution $\tilde u(x) = - 2(4-d)/x^2$ discussed in the previous
section in the
case $\beta=1$, is now (we set $\sigma=1$) 
\be
 \tilde u(x) = - (\frac{2(2- (d-2) \beta)}{\beta^2})^{1/\beta} 
\ee 
It implies 
(see Thm1$_\beta$ there) that the probability that the avalanche
started at $x=0$ remains within a sphere of radius $R$ is now $e^{f \tilde u(0)/R^{2/\beta}}$ 
where $u(0)=-\tilde u(0)$ is given in \cite[Prop. 3.5]{Iscoe88}. The upper critical
dimension is now $d=d_\beta=2 + \frac{2}{\beta}$ and the meeting
probability of the avalanche started at $x=0$ with a ball $B(x,\epsilon)$ 
now decays at large $|x|$ as $\sim |x|^{-2/\beta}$ for 
$d<d_\beta$, $\sim 1/(x^2 \log|x|)^{1/\beta}$ for $d=d_\beta$ and
$\sim \epsilon^{d-2-2/\beta}/|x|^{d-2}$ for $d>d_\beta$. 
The Thm 3$_\beta$ there also gives results about local and 
global extinction when starting from an 
extended initial measure (such as the Lebesgue measure).
More results about the regularity and irregularity of superprocesses 
with (1+$\beta$)-stable branching with $\alpha=2$ are obtained in 
\cite{MytnikSuperprocess2015,MytnikStable2016}.

There seems to be only a few results for the case of general $\alpha$, $\beta$. 
One case is about the regularity properties of $\rho_t$ fixed time $t$.
If $d<\alpha/\beta$, for any fixed $t$, $\rho_t$ is absolutely continuous w.r.t. the Lebesgue measure,
hence a density $\rho_t(dx)=\rho(x,t) dx$ can be defined, while for $d< \alpha + \frac{\alpha}{\beta}$
the weighted occupation time random measure $\int_0^t ds \rho_s$ is absolutely continuous
\cite{Fleischmann1988,Fleischmann2010}. 
The fixed time density $\rho(x,t)$ was studied in $d=1$ for $\beta=1$ and $\alpha>1$ and shown to be Holder continuous $(\alpha-1)/2$. 
\cite{Konno-Shiga88}. More general results for Holder regularity are obtained in \cite{Fleischmann2010}
and in \cite{HughesDensity2020}.
Other interesting results for general $\alpha,\beta,d$ can be found in \cite{StableBranchingInitialPoisson}
where an initial inhomogeneous Poisson measure was considered.

Interestingly, it  is shown in \cite{MytnikMultifractal2012} that in space dimension $d=1$, the density is multifractal for $\alpha>\beta+1$. The case of infinite mean $\beta <1$ and in particular the limit $\beta \to 0$ has also been studied in
\cite{FleischmannSturm} and more recently in \cite{MytnikInfiniteMean}. More limit theorems can be found in \cite{RenSongSun} and relations to random trees in \cite{LeGallReviewLevySnake}. Finally
generalizations for Ornstein-Uhlenbeck superprocesses have been studied \cite{SuperOUProcess}. 
\\

On the side of the LR elastic systems there are many works which study avalanches, most of them from experiments 
or numerics and with short range disorder (i.e. away from the mean-field theory). 
An important work is Ref. \cite{Zapperi} where the PDF of the size $S_c$ of the connected clusters was
found to be a power law $P(S_c) \sim S_c^{- \tau_c}$ with an exponent $\tau_c$ different from the 
exponent $\tau$ for the PDF, $P(S_c) \sim S^{- \tau}$, of the total avalanche size. A relation between these
two exponents was proposed there. This relation was reexamined more recently, and extended to other
cluster observables in \cite{LePriolPRL2}. These studies are in general beyond mean-field, but they
include the case $d=2 \alpha=1$, which corresponds to the "upper critical dimension" where mean-field
should apply. The LR-BFM was also studied in Chap 5. of \cite{LePriolThesis}, where one can find a review of
the physics literature on this topic. The long range instanton equation \eqref{instLR} was studied
there, as well as in \cite{ThieryShape}, and observables such as $\langle \dot u(x,t) \rangle_S$
and $\langle S(x)^p \rangle_S$ were computed. Finally, it is interesting to note that some interface
systems in nature are indeed at the upper critical dimension $d=d_c=2 \alpha$, 
so that the LR-BFM can be measured in experiments, see e.g. \cite{PLDDurin2016}
where the prediction for the global avalanche shape was tested. To test the LR-BFM
beyond the LR-ABBM model one would need, however, to measure quantities which are local in space
which is possible in principle.
\\

Clearly much more remains to be done on this topic, and at this stage there seems to be little overlap between the existing results from superprocesses and those from LR avalanches.

\section{Conclusion} 

In this review we have attempted to put side by side the mean-field theory of avalanches of elastic interfaces in physics, 
and the theory of the super Brownian motion (SBM) and superprocesses in probability. There appears to be an equivalence
between the Brownian force model (BFM), which was derived in the physics context from the more general field theory 
of interfaces based on the functional RG, and the stochastic equation describing the time evolution of the measures in the SBM. Both are driven by the so-called demographic noise, and are defined in the continuum as scaling limits. The BFM is defined as the scaling limit of discrete interface models (such as used in numerical simulations) near criticality when $m^2 \to 0$, with proper rescaling of space and time,
and describe the full statistics of the velocity field $\dot u(x,t)$ in an avalanche, also called activity. The SBM is obtained as a scaling limit of discrete branching processes, or space-embedded Galton-Watson trees and describe the time evolution of a measure of density $\rho(x,t)$. The observables in each system are in correspondence, as summarized in Table I, most notably
 $\dot u(x,t)$ corresponds to $\rho(x,t)$. Both systems satisfy the same duality relation, which maps the calculation of some observables to solving a non linear PDE, called the "instanton equation" in the BFM context. Using this equation many observables have been computed on both sides (independently, since both fields have developed separately). However for some other observables, the focus was different on each side, and one can hope to learn a lot from the comparison. Much focus in the SBM community has been about the regularity of the activity field, and of the local avalanche size $L_x=S(x)$, and the spatial structure and Hausdorff dimensions of their supports. This is complementary to the information about averaged quantities, such as the spatial shape, obtained in the BFM.
 
In the study of avalanches of elastic interfaces the BFM is only a starting point, and the outstanding question is how to construct in a more precise way the field theory in the continuum limit beyond the mean-field theory
(see Appendix \eqref{app:FRG}), and extract from it the information about the random geometry of the avalanches for short range disorder.
This is particularly challenging in the case of long range elasticity where the statistics of the clusters remains mysterious. As was shown in FRG studies \cite{LeDoussalWiese2012a,DobrinevskiLeDoussalWiese2014a,ThieryShape} to lowest order in an $\epsilon=2 \alpha-d$ expansion, it requires to solve the "instanton equation" in presence of a random static linear perturbation. Although this allowed to compute some quantities (such as the total size avalanche exponent $\tau$) it appears much more difficult to define and extract from it detailed geometrical information, e.g. about the clusters. One can hope that the wealth of information coming from the results on the SBM and the superprocesses will help in that respect, in particular to formulate the questions in a more mathematically precise way.  It is interesting to note that there is an equivalent description of the elastic systems near depinning (i.e. in the avalanche regime) in terms of sandpile models, and most notably in terms of reaction diffusion models for the activity field (the so-called directed percolation with a conserved field)
\cite{PLDKWStoch2015,JanssenDP} which is more general, but quite similar, to the BFM-SBM correspondence.

\bigskip

{\it Notations:} we use interchangeably $\partial_t u$ or $\dot{u}$ for time derivatives,
$dx$ and $d^d x$ for the volume element, 
$\delta(x)$, $\delta^d(x)$ for a delta function at the origin, $d_H$ and ${\rm dim}$ to denote the Hausdorff dimension. 
We use $\delta_0$ 
for a delta-measure at $x=0$ and $\delta_{ij}$ for Kronecker delta. For intervals, $(a,b)$ does not include enpoints,
$[a,b)$ includes $a$.

\subsection*{Acknowledgements.} This work started in 2014 at the
Oberwolfach program "Stochastic Analysis: Around the KPZ Universality Class".
I thank Martin Hairer and Jeremy Quastel for fruitful discussions there, 
and pointing out literature about the SBM. I also thank G. Ben Arous and C. Giardina for an
interesting
discussion, and Edwin Perkins and Leonid Mytnik for correspondence.
I thank C. Le Priol , A. Rosso and T. Thi\'ery for discussions and
collaborations on avalanches and X. Cao for careful reading of the manuscript.
This work was supported initially by PSL grant ANR-10-IDEX-0001-02-PSL,
then by ANR grant ANR-17-CE30-0027-01 RaMaTraF.


\begin{appendix}



\setcounter{equation}{0}
\setcounter{figure}{0}
\renewcommand{\theequation}{A\arabic{equation}}
\renewcommand{\thefigure}{A\arabic{figure}}

\section{Bienaym\'e-Galton-Watson and Feller processes, ABBM model } 
\label{app:GW} 

{\bf Bienaym\'e-Galton-Watson process}. In the BGW process $M_n$ is the number of individuals alive at generation $n$. Each independently gives rise to $k$ children with probability $p_k$ leading to $M_{n+1}=\sum_{j=1}^{M_n} k_j$ with i.i.d $k_j$. One has 
\be
\mathbb{E}( z^{M_{n+1}} | M_n ) = \Phi(z)^{M_n} ~ , ~ \mathbb{E}( e^{\lambda M_{n+1}} | M_n )  = e^{M_n \phi(\lambda)} ~ , ~ \ \Phi(z)= \sum_{k \geq 0} p_k z^k = e^{\phi(\lambda)}  
\ee
with $z=e^\lambda$, both formulations being convenient. 
\\

\noindent
{\bf Feller process}. The continuum limit goes as follows. Let us define the process $X_t$ for times $t \in \frac1{N V} \mathbb{N}$ 
\be
X_t = \frac{1}{N} M_{n= N V t}  
\ee 
in the large $N$ limit, and tune the first moment $\langle k \rangle$ of $p_k$ close to unity, while keeping the second 
cumulant $\langle k^2 \rangle^c$, 
fixed 
\be \label{philambda} 
\phi(\lambda) = \langle k \rangle \lambda + \frac{1}{2} \langle k^2 \rangle^c \lambda^2 + O(\lambda^3) 
\quad , \quad \langle k \rangle = 1 + \frac{\gamma}{N V} 
\ee 

In the continuum limit it goes to the stochastic process (with $X_t \geq 0$) 
\be \label{stoch0} 
d X_t = \sqrt{ \tilde b X_t} dB_t + \gamma X_t dt   \quad , \quad \tilde b = V \langle k^2 \rangle^c
\ee 
where $B_t$ is the standard Brownian motion.
Indeed consider its change from time $t$ to time $t+dt$ with $dt = \frac{1}{N V}$. The calculation
being identical for all $t$ let us choose for convenience $t=0$. One has 
\be 
\mathbb{E}( e^{\mu X_{dt}} ) - e^{\mu X_0} = \mathbb{E}( e^{\frac{\mu}{N} M_1} ) - e^{\mu X_0} 
= e^{M_0 \phi(\frac{\mu}{N}) } - e^{\mu X_0} 
= e^{N X_0 \phi(\frac{\mu}{N}) }  - e^{\mu X_0} 
\ee 
Using \eqref{philambda} and expanding at large $N$ one has
\be \label{dd} 
\mathbb{E}( e^{\mu X_{dt}} ) - e^{\mu X_0} = e^{\mu X_0}  ( 
e^{ \frac{\gamma}{N V} \mu X_0 + \frac{1}{2 N} \langle k^2 \rangle_c X_0 \mu^2 
+ o(\frac{1}{N}) } - 1) 
= (\gamma X_0 \mu  + \frac{1}{2} V \langle k^2 \rangle_c X_0 \mu^2 ) e^{\mu X_0} dt + o(1)
\ee
On the other hand for the stochastic equation \eqref{stoch0} one has using Ito's rule
\be \label{de} 
d (e^{\mu X_{t}}) = \mu e^{\mu X_{t}} dX_t + \frac{1}{2} \mu^2 \tilde b X_t e^{\mu X_{t}} dt 
\ee 
Taking the expectation value of \eqref{de} one obtains the same form as in \eqref{dd},
which allows to identify the two processes in the large $N$ limit. For more
details and rigorous statements see e.g. \cite{Feller51,Etheridge}. 

Consider now BESQ$^{\sf d}$ is the square of the ${\sf d}$-dimensional Brownian 
motion. By definition $Y_{\sf d}(t)= \sum_{i=1}^{\sf d} B_i(t)^2$ hence
\be 
d Y_{\sf d} = 2 \sum_i B_i dB_i + {\sf d} \, dt 
\ee 
Since the $B_i$ are independent Brownian motions, the sum $\sum_i B_i dB_i$ 
can be replaced by $$\sqrt{ \sum_i B_i^2 } dW = \sqrt{Y_{\sf d} } dW$$ where $W$ is
another Brownian motion and one has
\be 
d Y_{\sf d} = 2 \sqrt{Y_{\sf d} } dW + {\sf d} \, dt 
\ee 
which, up to the rescaling $X_t = \frac{\tilde b}{4} Y_{\sf d}(t)$ 
is the same stochastic equation as \eqref{stoch0} with $\gamma=0$ (critical case). In addition
there is a constant {\it immigration} term $d Y_{\sf d} = {\sf d} \, dt$, i.e. $d X_t =   \frac{\tilde b}{4} {\sf d} \, dt$ 
which in the BGW process corresponds to 
$M_{n+1}=\sum_{j=1}^{M_n} k_j + m_n$ with $m_n= \tilde b {\sf d} /(4 V)$. 
\\

Let us recall that, by contrast, the norm of the ${\sf d}$-dimensional Brownian motion 
started from the origin, $X_t = ||{\bf B}_t||$ is the 
Bessel({\sf d}) process, which obeys $dX_t = \frac{{\sf d}-1}{2} \frac{dt}{X_t} + dB_t$. 
\\

To obtain a solution to \eqref{stoch0} with non zero $\gamma$ from a solution with $\gamma=0$ 
one performs a time change $t \to \tau(t) = \frac{1}{\gamma}(1- e^{-\gamma t})$, with $d\tau= e^{- \gamma t} dt$. Denote 
$Z_t= e^{\gamma t} Y_{\sf d}(\tau(t))$ one has 
\be 
d Z_t = \gamma Z_t dt + e^{\gamma t} dY_{\sf d}(\tau) =
\gamma Z_t dt  + e^{\gamma t} ( 2 \sqrt{Y_{\sf d} } dW(\tau) + {\sf d} \, d\tau  ) 
= ({\sf d} + \gamma Z_t) dt + 2 \sqrt{Z_t } dB(t) 
\ee 
using that $dW(\tau) \equiv e^{- \gamma t/2} dB(t)$ where $B(t)$ is a Brownian in $t$. 
Upon rescaling the solution of  \eqref{stoch0} is $X_t = \frac{\tilde b}{4} Z_t$, 
as indicated in the text (where we set $\tilde b=4$).
\\

{\bf ABBM model}. The ABBM model describes the center of mass of an interface at time-dependent position $u(t)$
which is driven in a random medium by a spring of strength $m^2$ (also called the mass), i.e.
a parabolic potential, of center at position $w(t)$, which is imposed. The evolution equation is 
(setting $\eta=1$) 
\be \label{ABBM1} 
\partial_t u(t) = m^2 (w(t) - u(t)) + F(u(t))  
\ee 
In the ABBM model the random force lansdcape is Brownian, i.e. 
$F(u)= \sqrt{2 \sigma} B(u)$, where $B(u)$ is a Brownian motion in $u$.
Taking a derivative of 
\eqref{ABBM1} w.r.t. time, one obtains (with $\dot u=\partial_t u$) 
\be \label{ABBM2} 
d \dot u(t) = m^2 (\dot w(t) - \dot u(t) ) dt +  F'(u(t)) \dot u(t) dt
\ee
For forward driving $\dot w \geq 0$ the motion is forward, $\dot u \geq 0$. Hence one can
rewrite this equation as a stochastic equation for $\dot u\equiv \dot u(u)$ as a function of $u=u(t)$ itself (writing $dt=du/d\dot u$) 
\be \label{ABBM3} 
d \dot u = m^2 (\frac{\dot w(t)}{\dot u}  - 1) du + F'(u) du = m^2 (\frac{\dot w(t)}{\dot u}  - 1) du + \sqrt{2 \sigma} dB(u)  
\ee 
This equation is often studied by driving at fixed velocity $\dot w(t)=v$. Near the origin
$\dot u=0$ one can neglect the term $- m^2 du$ and the motion is identical to the Bessel process $X_u = ||{\bf B}_u||$,
with $\dot u(u)= \sqrt{2 \sigma} X_u$, in dimension ${\sf d}'= 1+ \frac{m^2 v}{\sigma}$. 
For $v>0$ and ${\sf d}' <2$ the velocity $\dot u$ thus vanishes infinitely often. One defines
the avalanche sizes $S=\int dt \dot u(t) = \int du$ between two consecutive vanishing of $\dot u$.
Their distribution is thus the same as the one for the return to the origin of the Brownian motion 
in dimension ${\sf d}'$ and 
one finds 
\be 
P(S) \sim S^{-\tau(v)} \quad , \quad \tau(v) = 2 - \frac{{\sf d}'}{2} =  \frac{3}{2} - \frac{m^2 v}{2 \sigma} 
\ee 
for $S \ll S_m=\sigma/m^4$. This holds for $v<v_c=\sigma/m^2$ such that $\tau(v_c)=1$. For larger driving velocity ${\sf v}$ one has ${\sf d}'>2$ leading eventually to an infinite avalanche (the velocity does not returns to zero). 
\\

{\it Remark}. This transition at $v=v_c$ is an interesting feature of the ABBM model. It can be translated to the SBM in presence of immigration: if the immigration rate is small enough there will be infinitely many extinctions, while if it is larger, after some time, there will be no more extinction. This holds in any space dimension since the total avalanche size (and total mass in the SBM) is described by the ABBM model (the Feller process). 
\\

Alternatively, starting again from \eqref{ABBM2}, and since $F'(u)$ is a white noise 
\be 
\mathbb{E} [ F'(u(t)) F'(u(t')) ] = \delta(u(t)-u(t')) = 2 \sigma \delta(t-t') \frac{1}{\dot u} 
\ee 
one obtains the stochastic equation for $\dot u = \dot u(t)$ as a function of time 
\be \label{ABBM4} 
d \dot u = m^2 (\dot w(t) - \dot u ) dt +   \sqrt{2 \sigma \dot u } \, dW(t) 
\ee
where $W(t)$ is a Brownian motion in time. Now this identifies with the Feller process 
with immigration, i.e. 
\be 
\dot u(t) = \frac{2 \sigma}{\tilde b} X_t = \frac{\sigma}{2} e^{\gamma t} Y_{\sf d}(\frac{1}{\gamma}(1-e^{-\gamma t}))
\quad , \quad \gamma = - m^2 \quad , \quad 
\frac{m^2 v}{\sigma} = \frac{{\sf d}}{2} 
\ee 
where we recall $Y_{\sf d}(t)={\bf B}_t^2$ is BESQ$^{\sf d}$, the square of the $d$-dimensional Brownian 
motion. Note that although both processes $\dot u(u)$ and $\dot u(t)$ are related to the Bessel process,
the relation is not identical (since the time change is itself random) and as a result the dimensions
${\sf d}$ and ${\sf d}'$ are different (one has $\frac{{\sf d}}{2} = {\sf d}'-1$, which coincide however 
for the critical case ${\sf d}={\sf d}'=2$). 
\\


\section{Brownian force model from the functional RG field theory} 
\label{app:FRG} 

The BFM can be seen as a generalisation of the ABBM model (discussed in the previous section) including
space. The ABBM model is recovered either as the BFM in space dimension $d=0$, or,
more interestingly, as the exact evolution of the center of mass of the interface within the BFM. 

Let us recall here in a nutshell how the BFM arises as the mean-field theory for the avalanches at depinning
in the realistic case of short range disorder. 
It is a good description at and above its upper-critical dimension $d \geq d_c$ and it is
a starting point for a dimensional expansion in $d=d_c - \epsilon$.

For short-range disorder, the disorder can still be taken Gaussian, and 
is parameterized by its correlator 
\be
\overline{F(u,x)F(u',x')}= \delta^d(x-x') \Delta_b(u-u') \label{defF2} 
\ee
where $\Delta_b(u)$ is a smooth short range function, the index $b$ means "bare".

In the functional RG field theory approach of depinning, the function $\Delta(u)$ is the fundamental object
(here we refer only to \cite{LeDoussalWiese2012a} where all previous references can be found). 
One first defines the renormalized disorder $\Delta_m(u)$ in a very natural way as follows. 
Consider uniform driving $w(x,t)=w(t)$, monotonous but infinitely slow, $\dot w(t) \to 0^+$. 
One can show from Middleton's theorems that the position of the center of mass of the interface
$\bar u(t)=\frac{1}{L^d} \int d^d x \, u(x,t)$ becomes (after a transient)
a uniquely defined function $u(w)$ of $w$ (i.e. $\bar u(t) \to u(w(t))$). The renormalized
disorder is then defined by the second cumulant
\be \label{renDelta} 
\Delta_m(u) = L^d \, \overline{ {\sf F}(w) {\sf F}(w') }^c
\ee
where ${\sf F}(w)= m^2 (u(w)-w)$ the total driving force exerted on the interface, and
$L^d$ the system volume. It depends on $m$, with
the "initial condition" that for $m =+\infty$, $\Delta_m(u)=\Delta_b(u)$.
The functional RG field theory allows to derive a flow equation for $\Delta_m(u)$
as a function of $m$. To summarize its results, one finds that for $d \leq d_c$
and as $m \to 0$ it flows (upon rescaling) to a fixed point function form,
which describes the system near criticality. This fixed point form has been
obtained in an expansion in $\epsilon=d-d_c$. Remarkably, this fixed point form 
is non-analytic at $u=0$, it has a cusp, i.e. $\Delta(u)= \Delta(0) + \Delta'(0^+) |u|+ 
\Delta''(0^+) \frac{u^2}{2} +O(|u|^3)$, i.e. the renormalized disorder is not smooth.
This cusp is fundamentally connected to
avalanches. Indeed, it is easy to show from the definition \eqref{renDelta} that
\be
S_m := \frac{\langle S^2 \rangle}{2 \langle S \rangle} = \frac{-\Delta_m'(0^+)}{m^4} 
\ee 
since the c.o.m. position function $u(w)$ defined above exhibits jumps, corresponding to
avalanches (we recall that $-\Delta_m'(0^+)>0$). 

To show how the BFM arises, let us start from the equation of motion \eqref{BFMpos1} and take the time derivative
\be
\eta \partial_t \dot u(x,t)= (\nabla_x^2 - m^2) \dot u(x,t) + \partial_t F\!\left(u(x,t),x\right) +  \dot f(x,t) 
\label{BFMpos12}
\ee

Let us examine the disorder term. Its two point correlation is
\bea
&& \overline{\partial_t F\!\left(u(x,t),x\right) \partial_{t'} F\!\left(u(x',t'),x'\right)} =
\partial_t \partial_{t'} \Delta(u(x,t)-u(x,t')) \delta^d(x-x') \nn \\
&& = \dot u(x,t) \partial_{t'} \Delta'(u(x,t)-u(x,t')) \delta^d(x-x')  \nn \\
&& =
 \dot u(x,t) \partial_{t'} [ \Delta'(0^+) {\rm sgn}(u(x,t)-u(x,t'))  \delta^d(x-x')
 + O( \Delta''(0^+)  ) ] \nonumber \\
 && = - 2 \Delta'(0^+) \dot u(x,t)   \delta(t-t') \delta^d(x-x') + O( \Delta''(0^+) ) \label{BFMFRG} 
\eea
If we discard the higher order terms $O( \Delta''(0^+) )$ this gives {\it exactly} the random
force which appears in the BFM evolution equation \eqref{BFMdef1}, with the
identification $\sigma=- \Delta'(0^+)$.

In obtaining the last line in \eqref{BFMFRG} 
we have used (i) $\Delta'(u)=\Delta'(0^+)+ u \Delta''(0^+) + \dots$, i.e. an expansion
of the cuspy form of the disorder correlator (ii) ${\rm sgn}(u(x,t)-u(x,t'))={\rm sgn}(t-t')$ 
according to Middleton theorems. In obtaining the BFM, we have further 
neglected the higher order terms in the expansion (i). This requires
an explanation.

All these steps can be fully justified within the field theory. We will not reproduce it here
but only give the main idea. We refer to \cite{LeDoussalWiese2012a} for details. 
First, one must use of the renormalized disorder
$\Delta(u)=\Delta_m(u)$ in \eqref{BFMFRG}, and in particular its cuspy fixed point form. If one were doing
naive perturbation theory, one would use $\Delta_b(u)$ which is analytic. That does not lead
to the BFM. However, in the field theory RG approach, all observables are computed in 
perturbation theory {\it in the renormalized disorder}, since it is $\Delta_m(u)$
which is small near the fixed point around $d=d_c$. In practice
this is achieved using the Martin-Siggia-Rose dynamical action, which amounts
to introduce response fields $\tilde u(x,t)$ to enforce the equation of motion
for $\dot u(x,t)$ and average over the disorder. 
From this action one then computes the effective action functional, $\Gamma[\dot u, \tilde u]$.
One finds that the BFM (i.e. the lowest order in the expansion in \eqref{BFMFRG})
arises a {\it summation of all tree diagrams} in perturbation theory
(it is actually an improved tree diagram summation, as it also contains
the loop diagrams which lead to the cusp, see discussion in
\cite{LeDoussalWiese2012a}, Section III.G). This is
the usual way in field theory to construct the mean-field theory, and is expected to
be valid to describe avalanches for $d \geq d_c$ (and valid, but with some logarithmic corrections in $m$ 
at $d=d_c$).

An important corrolary, obtained in \cite{LeDoussalWiese2012a},
is that to describe the deviations from the mean-field theory, one can simply 
restore the next order term $O( \Delta''(0^+) )$ in the expansion \eqref{BFMFRG}.
It can be treated systematically as a perturbation in an expansion in $\epsilon=d_c-d$. 
To lowest order, its effect can be incorporated simply in a modification of the
instanton equation \eqref{instanton} by adding to it the term $\chi(x) \tilde u(x,t)$.
Here $\chi(x)$ is a time independent Gaussian random
field of correlator $\overline{\chi(x) \chi(x')} \sim \Delta''(0^+) \delta^d(x-x') = O(\epsilon)$. 
It is a fictitious field however as its variance has a negative sign. Nevertheless,
upon formally averaging over it, one obtains the corresponding result to $O(\epsilon)$. As mentionned
in the text it has allowed to obtain a number of observables beyond mean field, to order $O(\epsilon)$.

A systematic approach can be implemented, and one can generalize the "instanton equation"
beyond mean-field \cite{PLDunpub}. It reads
\be \label{generalinstanton} 
\frac{\delta \Gamma[\dot u, \tilde u]}{\delta \dot u(x,t)}|_{\dot u(x,t) \to 0^+} = \lambda(x,t) 
\ee 
Together with the exact equation
\be 
\log G[\lambda,f] + \Gamma[\dot u, \tilde u] = \int dt \int d^dx ( \lambda(x,t) \dot u(x,t) + 
\dot f(x,t) \tilde u(x,t) ) 
\ee 
which for $\dot u(x,t) \to 0^+$ recovers \eqref{generating}
for the generating function (here written $G[\lambda,f]$ for an arbitrary driving).
This allows
to obtain information about avalanches beyond the BFM (it recovers 
\eqref{instanton} if specialized to the BFM). This requires evaluating the l.h.s. of 
\eqref{generalinstanton} which can be performed systematically
in perturbation theory, or using truncated exact RG methods.

\section{Relation between SBM duality and BFM }
\label{app:relation} 

Let us rewrite and compare the duality property of the SBM, Eqs. \eqref{duality1}, \eqref{instantonSBM}
and the formula for the generating function in the BFM, Eqs. \eqref{generating}  \eqref{instanton}.

For the SBM it reads, with $t \geq 0$
\bea
&& \mathbb{E} [ e^{- \int d^d x \phi(x) \rho(x,t)} ] = e^{- \int_x \rho(x,0) v(x,t) } \\
&& \partial_\tau v = D \nabla_x^2 v - \tilde b v^2 + \gamma v  \label{instSBM2} 
\quad, \quad 
v(x,t=0)=\phi(x) \geq 0 \nonumber
\eea 
For convenience we assume a density and denote $\rho_t(dx)=\rho(x,t) dx$, which is valid for $d=1$, 
but the idea is the same for $d>1$ (and provides a rigorous setting for the BFM). 

For the BFM (choosing $\eta=1$) we denote for convenience the time as $\tau \geq 0$, and it reads
\bea
&& \overline{e^{\int_{\tau \geq 0} d\tau \int d^d x \lambda(x,\tau) \dot u(x,\tau)}} = 
e^{ \int_{\tau \geq 0} d\tau \int d^d x \dot f(x,\tau)  \tilde u(x,\tau)} \\
&& \partial_\tau \tilde u + \nabla_x^2 \tilde u + \sigma \tilde u^2 - m^2 \tilde u = - \lambda \label{inst4} \\
&& \tilde u \leq 0 \quad, \quad \tilde u(x,\tau)=0 \quad \tau> \tau^* >0
\eea 
where the latter boundary condition holds for sources 
with bounded support in time, $\tau<\tau^*$, and the instanton 
equation \eqref{inst4} should be integrated backward in time.
We recall that $f(x,\tau)=m^2 w(x,\tau)$. 

We want to identify $\rho(x,t) = \dot u(x,t)$. Hence we should
specialize in the BFM the driving $f$ to be a kick, and the source $\lambda$ as
\bea
&& \dot f(x,\tau) = \rho(x,0) \delta(\tau) \\
&& \lambda(x,\tau)= - \phi(x) \delta(\tau-t) \label{source2} 
\eea
The kick guarantees from \eqref{BFMdef1} (with $\eta=1$ and since the velocity
vanishes for $\tau<0$) 
that the initial velocity of the BFM equals the initial density of the SBM,
$\dot u(x,t=0^+)=\rho(x,\tau=0)$. Since one integrates \eqref{inst4} backwards in time, \eqref{source2} 
leads to $\tilde u(x,\tau=t^-) = - \phi(x)$, as $\tilde u(x,\tau)=0$ for $\tau>t$ (from
the boundary condition $\tilde u(x,+\infty)=0$). To identify the
two systems one needs to choose
\be \label{reluv} 
\tilde u(x,\tau) = - v(x,t-\tau) \quad , \quad 0 \leq \tau \leq t 
\ee
so that $\tilde u(x,0)=- v(x,t)$ as needed. One then checks that \eqref{inst4}
and \eqref{instSBM2} agree under the relation \eqref{reluv} 
with the choice $D=1$ and $\tilde b=\sigma$.

\section{More on observables and results}
\label{app:observables} 

Here we review a few more observables and results, in complement to Section \ref{sec:observables}.
The first subsection are results from BFM studies, which we aim to translate into the SBM framework, and
reciprocally for the second subsection. Let us first recall the following useful correspondence:
\\

{\bf Driving: SBM versus BFM}. The SBM with initial measure $\rho_0(dx)=f(x) dx$ corresponds to driving 
by a kick in the BFM with $\dot f(x,t)=f(x) \delta(t)$ (here and below we set $\eta=1$).
In the BFM several driving protocols have been studied:

(i) the local kick $\dot f(x,t)= f \delta(x) \delta(t)$ which leads to 
$\dot u(x,0^+)=f \delta(x)$ hence in the SBM it corresponds to an initial measure proportional to a delta
function weight at $x=0$, $\rho_0=f \delta_{0}$. 

(ii) the global or uniform kick $\dot f(x,t)= F \delta(t)$ which leads to 
$\dot u(x,0^+)=F$, hence in the SBM it corresponds to an initial measure proportional to the Lebesgue measure
$\rho_0(dx)=F dx$. 

\subsection{More from BFM, towards SBM.} {\,}

\medskip

{\bf Total avalanche size and duration}. Consider again the BFM. In the original units, the PDF of the total
size $S$ of an avalanche after a kick, with $f=\int d^d x f(x)$, reads
\be  \label{PfS} 
P_f(S)=\frac{f}{2\sqrt{\pi} \sigma^{1/2} S^{3/2}} e^{- (m^2 S- f)^2/(4 \sigma S)}
\ee
Expressing $S$ in units of $S_m=\sigma/m^4$ and $f$ in units of $m^2 S_m$ it
gives the expression in the text. The critical case is obtained setting $-\gamma=m^2=0$ in \eqref{PfS}. 

Translated in the SBM context, Eq. \eqref{PfS} gives the PDF of the integrated super-Brownian
excursion (ISE) total mass, $S=L=\int_0^{+\infty} dt \rho_t(\mathbb{R}^d)$ (see Table I), 
as a function of the initial mass $f=\rho_0(\mathbb{R}^d)$. We recall the relation
between parameters $m^2=-\gamma$, $\sigma=\tilde b$, $D=2$. 

The avalanche duration, i.e. the extinction time $T$ of the SBM, has the following
cumulative distribution function (CDF)
\be 
P_f(T<t)= \exp\left( - \frac{m^2 f}{ \sigma (e^{m^2 t/\eta}-1) } \right)  \to_{m=0}  
\exp\left( - \frac{\eta f}{\sigma t}  \right)
\ee 

Of course, the space does not play much role in both these observables, and because of the additivity property of the SBM, 
this coincide with the result for the $d=0$ model, i.e to the corresponding
observables for the Feller process, equivalently for the ABBM model.
\\


{\bf Joint PDF of the local sizes $S(x)$, i.e. of the integrated local time $L_x$}.
In \cite{ThieryLeDoussalWiese2015} the joint PDF of the full field of local sizes $x \to S(x)$ was obtained. It reads
up to a normalisation
\bea \label{PSX} 
&& P[ \{S(x) \}] \propto \frac{\det M}{\prod_x S(x)^{1/2}} \exp\left( - \int d^d x \frac{ (\nabla_x^2 \, S(x) - m^2 S(x) + f(x))^2}{4 \sigma S(x)} \right) \\
&& M(x,y) =-  \frac{1}{m^2} (\nabla^2)_{x,y} + \delta^d(x-y) (1 + \frac{\nabla_x^2 \, S(x) - m^2 S(x) + f(x)}{m^2 S(x)} ) 
\eea 
where $(\nabla^2)_{x,y}$ is the $d$-dimensional Laplacian (these formula extend straightforwardly to more general, e.g. long range,
diffusion generator). We note that these formula contain a functional determinant.

The joint density of local sizes (for an infinitesimal kick at point $y$) was also obtained there, and has the simpler expression
\bea
&& \rho_y[ \{S(x) \}] = \partial_{f(y)} P[ \{S(x) \}] |_{\{f(x)=0 \}}   \\
&& = 
\frac{S(y)}{\prod_x S(x)^{1/2}} BC[\{S(x) \}] 
\exp\left( - \int d^d x \frac{ (\nabla_x^2 \, S(x) - m^2 S(x))^2}{4 \sigma S(x)} \right) \nn
\eea
where $BC$ depends on the boundary conditions at infinity (but can be set to unity here). 
The above formula were used to determine the most probable spatial profile (or shape) of the avalanche. 

In Ref \cite{ThieryLeDoussalWiese2015} these results were first established for a model where space is discretized,
with an arbitrary elastic matrix,
and then the continuum limit was obtained. 

In principle, provided a rigorous meaning can be given to these measures in the continuum, these results can be
transported to the SBM to describe the measure of the field of the total local times $L_x=S(x)$. We are not
aware of any such results in the SBM literature, and deriving it in that context could be a worthwile challenge. 
\\

{\bf PDF of local avalanche size $S(x)$ in the BFM, the SBM and the BBM}. The PDF of the local size $S(x)$ was computed for the BFM in $d=1$
\cite{LeDoussalWiese2012a,Delorme2016,PLDLong2021}. It corresponds to the
PDF of the time integrated local time $L_x=S(x)$ for the SBM, although we are not aware that it was computed explicitly in this context. It is obtained by Laplace inversion w.r.t. $-\lambda$ of 
\be  \label{LT0} 
\overline{e^{\lambda S(x)}} = e^{\int dy f(y) u^\lambda(y)} 
\ee 
where, in the massless case $u^\lambda(x)$ is the solution of $\tilde u''(x) - m^2 \tilde u(x)+ \sigma  \tilde u(x)^2 = - \lambda \delta(x)$. Of course Eq. \eqref{LT0} is well known in the SBM context, see e.g. \cite[(1.3)]{Hong5} and references
therein. In the massless case $m=0$, setting $\sigma=1$, it reads 
\be  \label{instdelta} 
u^\lambda(x) =  - \frac{6}{(|x|+x_\lambda)^2} \quad , \quad (x_\lambda)^3 = - \frac{24}{\lambda}
\ee

For a global kick, $f(x)=F$, the PDF of $S(x)$ is independent of $x$, so one can choose $x=0$ and study $S_0=S(0)$. The Laplace
inversion of \eqref{LT0} in that case gives \cite{LeDoussalWiese2012a}
\bea \label{mass1}
P_{f(x)=F}(S_0) = \frac{2 \times 3^{1/3} F e^{6 F m/\sigma}}{\sigma^{2/3} S_0^{4/3}} {\rm Ai}( \frac{3^{1/3} (S_0 m^2 + 2 F)}{\sigma^{2/3} S_0^{1/3}} ) 
\eea 
where ${\rm Ai}$ is the Airy function. The joint PDF $P_{f(x)=F}(S,S_0)$ for a uniform kick was obtained in \cite{Delorme2016}.

For a local kick at $x=0$, $f(x)=f \delta(x)$, the PDF of $S(x)$ for any $x$ was obtained in \cite{PLDLong2021}. 
Let us focus on $x=0$ for simplicity. In the massless case, from \eqref{LT0} and \eqref{instdelta}, it is given by Laplace inversion as (setting $\sigma=1$) 
\be 
P_{f \delta(x)}(S_0) = LT^{-1}_{-\lambda \to S_0} 
e^{- \frac{3^{1/3}}{2} (- \lambda)^{2/3} f }
=  \frac{f e^{-\frac{f^3}{36 S_0^2}}}{\sqrt[3]{3} S_0^{5/3}}  ( y  {\rm Ai}(y^2) - {\rm Ai}'(y^2) ) \quad , \quad y=\frac{f}{2 \times 3^{1/3} S_0^{2/3}} 
\ee 
The associated density, obtained for $f \to 0$, is $\rho_0(S_0) = \frac{1}{3^{2/3} \Gamma(\frac{1}{3}) S_0^{5/3}}$ and exhibits a decay exponent $5/3$. It should give the measure for $L_{x=0}$ under the canonical measure in the SBM. 
\\

Can one derive an analogous result in the BBM ? A related observable to $S(x)$ is the total time integrated density at $x$ 
of the BBM tree, equivalently the total local time at $x$ , which we will denote $L^{\rm BBM}_x$, dropping the subscript when no confusion is possible
\be 
L^{\rm BBM}_y =\int_0^{+\infty} dt \sum_{j=1}^{M(t)} \delta((x_j(t)-y) 
\ee 
The method to compute the PDF of this quantity is given in the text in Eqs.
\eqref{instBBM1}, \eqref{instBBM2} setting $\Lambda(x,t)=\lambda \delta(x-y)$
leading to
\be  
\mathbb{E}_{\{x_j(0)\}}(e^{ \lambda L^{\rm BBM}_y })  
= e^{ \sum_{j=1}^{M(0)} \ln \tilde g(x_j(0),0) }
\ee
where $\tilde g(x,\tau)$ solves the backward equation
\bea 
- \partial_\tau \tilde g = \nabla_x^2 \tilde g + V ( \Phi(\tilde g) - 
\tilde g) + \lambda \delta(x-y)  \tilde g 
\eea
which one solves backward in $\tau$ starting from $\tilde g(x,\tau=+\infty)=1$ down to
to $\tau=0$. Let us focus for instance on binary branching $V (\Phi(z)-z) = b z^2 + a - (b+a) z$
in the critical case $a=b$. One finds that the solution is
\bea \label{cubic} 
\tilde g(x) = 1 - \frac{6}{b (|x-y| + B_\lambda)^2} 
\quad , \quad b B_\lambda^3 - 6 B_\lambda = - \frac{24}{\lambda} 
\eea 
The final result for the Laplace transform of the PDF $P(L_y)$ of $L_y=L_y^{\rm BBM}$ is thus
\be 
\int dL_y P(L_y) e^{\lambda L_y} = \prod_{j=1}^{M(0)} 
(1 - \frac{6}{b (|x_j(0)-y| + B_\lambda)^2} )
\ee 
where $B_\lambda$ is solution of the cubic equation \eqref{cubic}.
For instance, for a single individual initially at $x=0$ at $t=0$ one can inverse Laplace
at least formally (with manipulations as in \cite{Delorme2016,PLDLong2021})
\be L_0 P(L_0)= 
- \int \frac{d\lambda}{2 i \pi} e^{- \lambda L_0} 
\partial_\lambda (\frac{6}{b B_\lambda^2})  
= \frac{12}{b} \int \frac{dB}{2 i \pi B^3} e^{ \frac{24}{B (b B^2-6)} L_0} 
= - \frac{12}{b} \int dv e^{ - \frac{4}{v} \sqrt{v + b/6} L_0} 
\ee
upon changes of variables from $\lambda \to B$ and $B \to v=\frac{1}{B^2}-\frac{b}{6}$, and proper choices of the contours. We have not 
attempted to specify further and compute these contour integrals. 

We note that the expressions for the BBM are reminiscent of those for the BFM and the SBM. One 
can again go from one to the others by considering $M(0) \sim N$, $b \sim N$, 
$S(x)= \frac{1}{N} L^{\rm BBM}(y)$, and taking $N$ large. 
\\

\subsection{More from SBM, towards BFM} 

Let us now return to results from the SBM, and attempt to translate them in
the language of avalanches.
\\

{\bf Some linear observables}. Consider an avalanche with a seed at $x_0$ conditioned to have reached
the unit ball centered on the origin, i.e. under $P_{x_0}(.|S(B(0,1))>0)$,
and consider the linear observable $S_\phi=\int d^d x S(x) \phi(x)$.
It was shown in \cite[Thm1] {LegallOccupation} that for a large avalanche, i.e. $x_0 \to +\infty$, the PDF's of 
$S_\phi/|x_0|^{4-d}$ for $d \leq 4$ and of $S_\phi/(\log |x_0|)$ for $d = 4$
reach some limits, which in $d=4$ is an exponential distribution. 

The PDF of some linear observables (with some choice for $\phi(x)$) 
were also studied and could be explicitly computed in the BFM in 
\cite[SecVIIIB]{PLDLong2021}.
\\


{\bf Exit measures}.  One can generalize the static instanton equation Eq. \eqref{inst2n} to a domain $D$ and
its boundary $\Gamma=\partial D$ as follows
\be \label{inst2n2} 
\nabla_x^2 \tilde u(x)  + \tilde u(x)^2 = 0 \quad , \quad \tilde u(x)|_{x \in \Gamma}= - g(x)
\ee
where $g(x)$ is a given positive function. For $g(x)=+\infty$ solving this equation
gives the probability \eqref{prob0} that
$\Gamma$ is never reached by an avalanche started inside $D$. 
For general $g(x)$ one defines the exit measure $\rho_D$ 
on $\partial D$ as the mass started at $\rho_0$ which is stopped at
the instant it leaves $D$. Then one has, see \cite[1.11]{Hong5} from
\cite[Tm 6.1.2]{LegallCourse}
\be 
\mathbb{E}_{\rho_0} \bigg[ 
\exp\left( - \langle \rho_D , g \rangle \right)  \bigg] = \exp( \langle \rho_0 , \tilde u \rangle) 
\ee
This allows in principle to quantify how much "activity" in an avalanche has spread outside a
given domain. 
\\

%

%
%
%

{\bf Uniform kick: global and local extinction}. A standard type of driving for an interface is to perform a spatially uniform kick 
$\dot f(x,t)= F \delta(t)$. The total force, and thus the total avalanche size, diverges 
with the size of the interface. For an infinite interface it corresponds to starting the SBM with the uniform
Lebesgue measure $\rho(x,0)=1$. In that case, there is a mathematically subtle discussion about
whether there is extinction, local or global \cite{Iscoe88,sawyer79}, that is a finite duration
for the avalanche activity. In fact, Dawson \cite{Dawson77} 
states that there is an invariant measure $\rho_\infty$ for $d \geq 3$.
Qualitatively the descendent of any ball vanish but because of the transience of the Brownian motion, there is a steady stream of particles replacing them. Hence there is a limit of $\lim_{t \to +\infty} \rho_t(G)$ for any bounded open set $G$
but that this limit is zero for $d\leq 2$. 
In space dimention $d=1$ it was shown \cite[Thm 3]{Iscoe88} that for each $R$, there is a $T_R$
such that $\rho_t([-R,R])=0$ for all $t> T_R$ with $P(T_R > t) \sim t^{-1/2}$.
So in $d=1$ there is local extinction, as also shown for the BBM in \cite{sawyer79}. Indeed
it is claimed in \cite{sawyer79} that for the critical case with initial uniform measure, in $d \leq 2$ clumps form with 
$\mathbb{E}[\rho_t(K)] = c \, {\rm vol}(K)$ but ${\rm Var}[\rho_t(K)] \to \infty$. Regions between clumps
tend to become empty, ${\rm Prob}(\rho_t(K) >0) \to 0$ as $t \to + \infty$ for bounded sets $K$ and
$d \leq 2$. Bounded open sets $K$ can either be impersistent (they become and remain empty)
or persistent (they continue to have visitors
indefinitely but perhaps more and more infrequently). In Ref.
\cite{sawyer79} it is claimed that they are persistent in $d=2$ (although $P(\rho_t(K) >0) \to 0$) 
and impersistent in $d=1$. This is obtained from whether 
$\int d^dx p(x)$ converges or not, where $p(x)$ is the probability to reach the neighborhood
of a point 
distance $x$, and $p(x) \sim 1/x^2$ at large $x$. 
\\


{\bf Local time and occupation time measure}. The local time (or occupation time) of the SBM is a random measure
which has a continuous density w.r.t. the Lebesgue measure for $d \leq 3$ \cite{Sugitani1989}.
This implies that the SBM hits points for $d \leq 3$ and does not hit sets which are "too small"
for $d \geq 4$ \cite[Sec. 6.5]{Etheridge}. Indeed for $d \leq 3$ 
there is a random function $L_x^t$ such that $\int_0^t ds \langle \rho_s , \phi \rangle = \int d^d x \phi(x) L_x^t$. The total local time, $L_x=L_x^\infty$, is thus the analog of $S(x)$ in the BFM. More generally, for finite initial measures, since the 
extinction time of the SBM is finite almost surely, one can define the total occupation time measure as 
$I(A)= \int_0^{+\infty} ds \rho_s(A)$. For $d \leq 3$ this measure has a density, which is $L_x$. 
\\

{\bf Behavior near the seed}. 
A question is how does $L_x$ behaves near the seed (assumed here to be $x_s=0$). For an 
initial condition $\delta_0$ (which corresponds to a local kick \eqref{xs}) it was shown 
in \cite{Hong2,Hong3} that for $d=3$, one has for any $0< t < +\infty$ 
\be
L_t^x \simeq_{x \to 0} \frac{1}{2 \pi |x|} + (\frac{1}{2 \pi^2} \log 1/|x|)^{1/2} \, \omega 
\ee
$\omega$ is a standard normal random variable independent of $\rho_t$. In $d=2$ there is a
similar result, with a logarithmic divergence as $x \to 0$. This is shown
by proving (Thm 1.9) that the solution to the instanton equation in $d=3$ has the form 
$\tilde u^\lambda(x) \simeq \frac{\lambda}{2 \pi |x|} - \frac{\lambda^2}{4 \pi^2}  \log(1/|x|)$. 

Singularities near the seed where also found in the calculation of the average shape at fixed total size, $\langle S(x) \rangle_S$, for avalanches in and beyond mean-field \cite{ThieryShape}. 
\\


{\bf Behavior near the boundary of an avalanche}.  An interesting question is how does $S(x)$ behaves near the boundary of an avalanche.
It is conjectured for the SBM, see  \cite{Hong1,MytnikBoundary},
that $L_x$ is Holder continuous of index $4-d-\eta$ for any $\eta>0$ 
near the zero set of $L_x$. In $d=1$ this conjecture was proved in \cite{Hong1} 
where it was shown that, for an initial condition $\delta_0$ (seed at $x=0$),
for any $\gamma>3$ there is a (random) $\delta>0$ such that a.s.
\be 
L_x \geq 2^{-\gamma/2} (R-x)^\gamma \quad , \quad R-\delta < x < R
\ee 
where $R$ is the right edge of the avalanche (we recall that the support of
the avalanche in $d=1$ was recalled to be an open interval $(L,R)$ \cite[Thm 1.7]{MytnikBoundary}). 
This seems consistent with the result for the BFM for $d=1$ mentioned in the text, 
i.e. that $S(x) \simeq (R-x)^3 \sigma$, where the PDF of the random variable
$\sigma$ was obtained in \cite{PLDLong2021}. The above conjecture for the SBM then
seems to agree qualitatively with a similar and more general conjecture, that the exponent 
for the vanishing of $S(x)$ near a boundary should be 
$\zeta$, the roughness exponent, equal for the BFM to $\zeta_{\rm BFM}=4-d$.

The question of the behavior of $L_x$ near the boundary is also related to the
question of the fractal dimension of the boundary, 
and we refer to \cite{MytnikBoundary,MytnikBoundary2018}
for all subtelties and recent results. 
\\

{\bf Behavior near the instantaneous boundary}. One can also study
the set of "active" points at a given time $t$ in an avalanche, 
such that $\dot u(x,t)>0$. In particular one can ask about the
boundary of this set, i.e. where $\dot u(x,t)$ is zero, but where
$\dot u(x,t)>0$ infinitesimally closeby. In the SBM the analog
quantity to $\dot u(x,t)$ is the density $\rho(x,t)$, which in
$d=1$ is well defined and continuous. This boundary set, called $BZ_t = \partial\{x : \rho(x,t) > 0\}$,
was studied in \cite{Mytnik1d,Perkins2018} in $d=1$. Usually $\rho(x,t)$ is Holder $1/2-\epsilon$,
but near the boundary it is more regular since the noise is mollified and is 
expected to be Holder $1-\epsilon$. The authors of \cite{Mytnik1d} prove that the Hausdorff dimension of the boundary is
\be 
0<{\rm dim} ( BZ_t ) = 2 - 2 \lambda_0 <1
\ee
and that 
\bea
{\rm Prob}(0<\rho(x,t)<a) \sim t^{-1/2-\lambda_0} a^{2 \lambda_0-1} 
\eea 
for all $0 < a < \sqrt{t}$. The non trivial exponent $\lambda_0$ obeys 
$1/2<\lambda_0<1$ and is equal to minus the largest eigenvalue of a killed
Ornstein Uhlenbeck process. Let us give the idea in a nutshell \cite{Mytnik1d}.
This remarkable result is obtained
by studying the solution to the time dependent instanton equation (in the BFM language)
$\tilde u^\lambda(x',t')$ with a source $\lambda \delta(x'-x)\delta(t'-t)$
in the limit $\lambda \to +\infty$. They show that the large $\lambda$
perturbation theory, 
$\tilde u^{+\infty}- \tilde u^\lambda \sim \lambda^{-\alpha}$,
determines the small $a$ behavior of ${\rm Prob}(0<\rho(x,t)<a) \sim a^\alpha$. 
Given that $\tilde u^{+\infty}(x,t)$ takes the scaling form $t^{-1} F(x/\sqrt{t})$,
where $F$ obeys a non-linear ODE, this perturbation theory is related 
to a killed Ornstein Uhlenbeck process of generator $h''/2 - (x/2) h' - F(x) h$.

\section{BBM calculations }
\label{app:BBMcalc} 

Here we recall some standard methods for derivation of results
in the BBM and the correspondence with the BFM.
\\

{\bf BBM: expectation of test functions}.
Consider as in the text the BMM where the particles alive at time $t$ have positions $x_j(t)$,
$j=1,.. M(t)$, with diffusion coefficient $D$ (so that $dx_i^2=2 D dt$). They die 
at rate $V$ and if so they immediately each produce $k \geq 0$ offsprings with probabilities $p_k$,
where $\Phi(z)=\sum_{k \geq 0} p_k z^k$. Note that the actual death rate is thus $V p_0$. 

Let us recall how to obtain heuristically (see \cite{McKean75,Etheridge} for rigorous derivations) 
the time evolution of expectation values of 
the type
\be 
g(x,t)= \mathbb{\mathbb{E}}_{\{x\}} \left( \prod_{j=1}^{M(t)} h(x_j(t)) \right)  \label{defg2} 
\ee 
for any test function $h$, where at $t=0$ there is a single particle alive, $M(0)=1$, at position $x_1(0)=x$. 
One decompose $[0,t+dt]$ in $[0,dt]$ and $[dt,t+dt]$. 
In the first interval $[0,dt]$ the initial first particle either:

(i) dies and produces $k$ offsprings, with probability $V p_k dt$ and 
in that case $g(x,t+dt)= g(x,t)^k$, by time translational
invariance, since $[dt,t+dt[$ has length $t$.

(ii) does not die and diffuses by $dx$ with $\langle dx^2 \rangle =2 D dt$.
This happens with probability $1- V dt$.
Then $g(x,t+dt)= \langle g(x+dx,t) \rangle_{dx} = g(x,t) + D \partial_x^2 g(x,t) dt + o(dt)$. 

Putting together we have 
\be
g(x,t+dt)= (1- V dt) (g(x,t) + D \partial_x^2 g(x,t) dt) + \sum_{k \geq 0} V p_k g(x,t)^k dt 
\ee 
which leads to Eq. \eqref{kpp1} in the text, namely 
\be  \label{kpp1n} 
 \partial_t g = D \nabla_x^2 g + V( \Phi(g)-g ) 
\ee
with, by definition, the initial condition $g(x,t=0)=h(x)$.

If there is more than one particle alive at $t=0$ the expectation value in 
\eqref{defg2} is simply equal to $\prod_{j=1}^{M(0)} g(x_j(0),t)$ (with the same function $g(x,t)$ as
above) since the $M(0)$ trees are independent. 
\\

Note that if one defines $\hat g_x(y,t)= \mathbb{\mathbb{E}}_{\{x\}} \left( \prod_{j=1}^{M(t)} \hat h(y-x_j(t)) \right)$
then, using translational invariance, one has $\hat g_x(y,t)=g(x-y,t)$ with a test function $h(z)=\hat h(-z)$. Hence
$\hat g_x(y,t)$ satisfies now $\partial_t \hat g = D \nabla_y^2 \hat g + V( \Phi(\hat g)-\hat g )$
with initial condition $\hat g_x(y,0)=\hat h(y-x)$. This is used below with the choice $\hat h(z)=\theta(z)$. 
\\

{\bf Distribution of rightmost particle at time $t$ in $d=1$}. Consider the BBM in $d=1$ 
with a single particle at position $x_1(0)=x$ at $t=0$, 
and denote $x_{\rm max}(t)=\max_{1 \leq j \leq M(t)}  x_j(t)$. One defines its CDF as
\be \label{Pxdef} 
P_x(y,t) = {\rm Prob}(x_{\max}(t)<y | x_1(0)=x) 
\ee 
more properly defined as the probability that the interval $[y,+\infty[$ is empty of particles at
time $t$ (so that if $M(t)=0$ this probability is one by convention). 
To compute this probability one can choose $h(z)=\theta(y-z)$
in Eq. \eqref{defg}. Then one has
\be 
P_x(y,t) = g_y(x,t)
\ee 
where 
$g_y(x,t)$ is the solution of \eqref{kpp1} with $g_y(x,t=0)= \theta(y-x)$.
\\

Another standard way to compute $P_x(y,t)$ is to derive an equation with respect to $y$. 
Consider binary branching.
Again one decompose $[0,t+dt]$ in $[0,dt]$ and $[dt,t+dt]$. 
In the first interval $[0,dt]$ the initial first particle either branches,
with rate $b$, or diffuses by $dx$ with $\langle dx^2 \rangle =2 D dt$, or dies. 
If it branches one has $P_x(y,t+ dt) = P_x(y,t)^2$, by time translational
invariance, since $[dt,t+dt[$ has length $t$. It if diffuses 
then $P_x(y,t+ dt) = \langle P_{x+dx}(y,t) \rangle_{dx} =
\langle P_x(y-dx,t) \rangle_{dx}$ where we used translational
invariance. Putting together we have 
\be
P_x(y,t+ dt) = \langle P_x(y-dx,t) \rangle_{dx} (1- (a+b) dt) + P_x(y,t)^2  \, b dt + a dt
\ee
leading to
\be \label{KPPPx} 
\partial_t P_x = D \partial_y^2 P_x + b P_x^2 - (b+a) P_x + a  \quad, \quad P_x(y,t=0)=\theta(y-x)
\ee 
which obeys also $P_x(+\infty,t=0)=1$.

Using translational invariance and parity, one can exchange the roles of the starting point $x_1(0)$ and $y$,
and obtain equivalently that 
$G(x,t)={\rm Prob}(x_{\max}(t)<x|x_1(0)=0)$ satisfies also \eqref{kpp1} with initial condition 
$G(x,0)=\theta(x)$. Indeed, for $x<y$ one has 
$g_y(x,t)={\rm Prob}(x_{\max}(t)<y | x_1(0)=x) = {\rm Prob}(x_{\max}(t)<y-x|x_1(0)=0)
= g_{y-x}(0,t)$. Hence $G(X,t)={\rm Prob}(x_{\max}(t)<X|x_1(0)=0)=g_X(0,t)=g_0(-X,t)$
satisfies \eqref{kpp1} with $G(X,0)=\theta(X)$. 


Consider the equation \eqref{KPPPx} for the CDF of the maximum defined in \eqref{Pxdef}.
In the super-critical case $b>a$ the the population grows exponentially and the 
large time asymptotics is described by a traveling wave.
In the critical and subcritical cases the population
decreases and at large time $P_x(y,t)$ converges to unity.  For
instance in the critical case $b=a$ one has that $\hat P(y,t)=1-P_0(y,t)$
satisfies $\partial_t \hat P = D \partial_y^2 \hat P - b \hat P^2$ with
$\hat P(y,0)=\theta(-y)$. Its uniform mode decreases to zero as $1/(b t)$ 
which is the probability that $M(t)=0$. 
\\

The analogous observable in the BFM is the probability that at time $t$ one has $\dot u(z,t)=0$, for all $z>y$ (i.e. no activity). 
It is given by $P^{\rm BFM}_x(y,t) = e^{- f \tilde u(x,0)}$ where $\tilde u(x,\tau)$ is the time-dependent solution of the instanton equation,
$\partial_\tau \tilde u + \partial_x^2 \tilde u - m^2 \tilde u + \sigma \tilde u^2 = 0$, 
with $\tilde u(z,\tau=t) = -\infty$ for $z>y$ and $\tilde u(z,\tau=t) = 0$ for $z<y$. 
\\


{\bf Distribution of the maximum up to time $t$ in $d=1$}.
Let us consider the probability, for a BBM 
\be
Q(y,t) = {\rm Prob}(X_m(t)<y) \quad , \quad X_m(t) = \max_{0 \leq s \leq t} \max_{1 \leq j \leq M(s)}  x_j(s) 
\ee
One can obtain it using Eqs. \eqref{instBBM1}, \eqref{instBBM2} by choosing
$\Lambda(x,\tau)=-\infty$ for $x> y$ and $0<\tau<t$ and 
$\Lambda(x,\tau)=0$ elsewhere. Thus the result is
\be 
Q(y,t)=\prod_{j=1}^{M(0)} \tilde g(x_j(0),0)
\ee 
where the function $\tilde g(x,\tau)$ is the solution of
\be  \label{instBBM3nn} 
- \partial_\tau \tilde g = \nabla_x^2 \tilde g + V ( \Phi(\tilde g) - 
\tilde g) \quad , \quad \tilde g(x>y,\tau)=0  \quad , \quad t^- > \tau >0 \quad ,\quad 
\tilde g(x,\tau=t^-)=\theta(y-x)
\ee
which is solved backward in time from $\tau=t^-$ to $\tau=0$. Note that for any time
$t^- > \tau >0$ one has also $\tilde g(-\infty,\tau)=1$. 
\\

Consider now $M(0)=1$ and denote $Q_x(y,t) = {\rm Prob}(X_m(t)<y | x_1(0)=x)$. 
In the critical and subcritical cases it reaches a non-zero stationary limit as $t \to +\infty$.
This limit is $q(x)=\lim_{t \to +\infty} Q_x(y,t)$, the probability that
no offsprings of an individual starting at $x$ has visited the level $y$ discussed in the text. It is then
given by the stationary solution of \eqref{instBBM3nn} 
\bea \label{instBBM4n} 
0 = D \nabla_x^2 q + V ( \Phi(q) - q)  \quad , \quad q(x=y) =0 
\eea
with $q(x) \to 1$ at $x \to - \infty$. The explicit solutions were recalled in the text. 
The relaxation to stationary state is as $1/t^2$ \cite{SatyaBBM4}. 
\\

A more standard way to perform the calculation is to note that $Q_x(y,t)$ is the solution for $y \geq 0$
of 
\be \label{aga2} 
\partial_t Q_x = D \partial_{y}^2 Q_x + V ( \Phi(Q_x) - Q_x)  \quad , \quad 
Q_x(x,t)=0 \quad , \quad Q_x(+\infty,t)=1 \quad , \quad Q_x(y,0)=\theta(y-x) 
\ee 
The two methods are consistent. Indeed let us denote $\tilde g_{yt}(x,\tau)$ the solution of
\eqref{instBBM3nn}. By time and space translational invariance it is a function of $t-\tau$ and $y-x$,
hence it is of the form $\tilde g_{yt}(x,\tau)=Q_0(y-x,t-\tau)$, and one can check
that $Q_0$ is precisely the solution of \eqref{aga2} for $x=0$. Since $Q_0(y-x,t-\tau)=Q_x(y,t-\tau)$ 
from space translational invariance, we see that $\tilde g_{yt}(x,\tau=0)=Q_x(y,t)$, as claimed. 
\\

\section{Multi-time observables in the BBM and the limit to the BFM}
\label{app:multitime} 

Here we give more details about the formula \eqref{instBBM1}, \eqref{instBBM2}
in the text, about weighted occupation
time,
which are the analog for the BBM 
of the "instanton equation" of the BFM, and recover its BFM limit.

\subsection{Two-time observables, a simple example} 

To start with a simple example let us consider the BBM with binary branching with death,  i.e. all $p_k=0$ except 
$p_2$ and $p_0$ with $p_0+p_2=1$
\be 
V (\Phi(z)-z) = V (p_2  z^2 + p_0-z) =  b z^2 +a - (b+a) z \quad , \quad p_2=\frac{b}{b+a} \quad , \quad p_0=\frac{d}{b+a} 
\quad , \quad V=b+a
\ee
with $\Phi(1)=1$, $\Phi'(1)= 1 + (b-a)/(b+a)$.
\\

Let us study the statistics of $M(t)$ the number of particles alive at time $t$,
and start with the one-time probabilities $P(M(t)=n)$. For this we choose a 
test function which is simply a constant, $h(x)=h$, $0<h<1$ and consider
\bea
g_{h}(t) = \mathbb{E}_1(h^{M(t)}) =\sum_{n \geq 0}  P(M(t)=n) \, h^n 
\eea 
where the subscript $1$ means that $M(0)=1$. According to \eqref{kpp1} the function $g(t)=g_h(t)$ is the solution of
\be  \label{KPPsimple} 
 \partial_t g = b g^2 +a - (b+a) g \quad, \quad g(t=0)=h
\ee
The explicit solution is
\be
g_h(t) = \frac{(b h - a) e^{t (a-b)} +a (1-h)}{(b h - a) e^{t (a-b)}+b(1-h)}
\ee
and for the critical case $a=b$
\be
g_h(t) =  \frac{b (1-h) t+h}{b (1-h) t+1} = \frac{b t}{1+ b t} + \sum_{n \geq 1} 
\frac{(b t)^{n-1}}{(b t+1)^{n+1}} h^n
\ee
from which one reads the probabilities $P(M(t)=n)$, in particular the survival probability
$P(M(t)>0) =1/(1+ b t)$ in the critical case. 

Note that for a more general initial condition, $M(t)=M(0)$, there are $M(0)$ independent trees, hence
\be 
\mathbb{E}_{M(0)}(h^{M(t)}) = (\mathbb{E}_{1}(h^{M(t)}) )^{M(0)} = g_h(t)^{M(0)} 
\ee
\\

We now want to study the joint probabilities of $M(t_1)$ and $M(t_2)$, for $t_2>t_1$. We thus define
for $h_1,h_2 \in [0,1]$ the generating function
\be
 g_{h_1,h_2}(t_1,t_2) = \mathbb{E}_1(h_2^{M(t_2)} h_1^{M(t_1)}) = 
 \sum_{n_1,n_2 \geq 0}  P(M(t_1)=n_1,M(t_2)=n_2) \, h_1^{n_1}  h_2^{n_2}
\ee
It can be written in terms of $g_h$ itself as
\bea
g_{h_1,h_2}(t_1,t_2) &=&  \mathbb{E}_1 [ \mathbb{E}_{M(t_1)} (h_2^{M(t_2)} )  h_1^{M(t_1)} ] \\
&=& \mathbb{E}_1 [g_{h_2}(t_2-t_1)^{M(t_1)}  h_1^{M(t_1)} ] 
= g_{h_1 g_{h_2}(t_2-t_1)}(t_1) \label{g2time} 
\eea 
Note the properties
\bea
&& g_{h_1,h_2}(t_1,t_1) = g_{h_1 h_2}(t_1)  \\
&& g_{h_1=0,h_2}(t_1,t_2) = g_{h_1=0}(t_1) = P(M(t_1)=0) \\
&& \partial_{h_1}|_{h_1=0} g_{h_1,h_2}(t_1,t_2) = P(M(t_1)=1) g_{h_2}(t_2-t_1) 
\eea

In the critical case $b=a$, the generating function is a function of $b t_1$ and $b t_2$, so we can set $b=1$ with no loss of generality, and we obtain from \label{g2time}
\bea
g_{h_1,h_2}(t_1,t_2) = \frac{h_1 \left(t_1-1\right) \left(h_2 \left(t_1-t_2+1\right)-t_1+t_2\right)+t_1
   \left(-h_2 t_1+\left(h_2-1\right) t_2+t_1-1\right)}{\left(h_1-1\right) t_1
   \left(\left(h_2-1\right) t_1+h_2\right)+t_2 \left(h_2-\left(h_1-1\right)
   \left(h_2-1\right) t_1\right)-t_2-1}
\eea
From this one extracts the joint probabilities, for instance
\bea
&& P(M(t_1)=0,M(t_2)=0) = \frac{t_1}{1+t_1} \quad , \quad 
 P(M(t_1)=0,M(t_2)> 0) = 0 \\
&& P(M(t_1)=1,M(t_2)=M)  = \frac{1}{(1+t_1)^2} p_{t_2-t_1}(M) \\
&& P(M(t_1)=M>0,M(t_2)=0)  = t_1^{M-1} \left(t_1+1\right){}^{-M-1} \left(1+t_2-t_1 \right){}^{-M}
   \left(t_2-t_1\right){}^M
\eea 

\subsection{Multi-time observables, a simple example} 

We want to generalize this to arbitrary number of times. We see that Eq. \label{g2time} amounts to
solve the equation \eqref{KPPsimple} for the last time slice $t_2-t_1$ starting from $h_2$,
then multiply the solution by $h_1$, then solve for the time slice $t_1$. Thus 
to calculate for $0<t_1<\dots<t_n$
\be \label{prod1} 
g(t_n) = \mathbb{E}_1(f_n^{M(t_n)} \dots f_2^{M(t_2)} f_1^{M(t_1)}) 
\ee
where we indicate only the dependence in the last time $t_n$, one defines
\be
\tilde g(\tau) = g(t_n-\tau)  
\ee
which satisfies the backward equation
\be \label{sourceshj} 
- \partial_\tau \log \tilde g(\tau) = b \tilde g(\tau) + \frac{a}{\tilde g(\tau)} - (b+a) + \sum_{j=1}^n \ln h_j \delta(\tau-t_j)
\ee
which one solves backward in $\tau$, starting from $\tilde g(\tau=t_n^+)=1$ down to
to $\tau=0$, and then one retrieve $g(t_n) = \tilde g(0)$.

For instance, for $n=2$, let us solve backward starting from $\tilde g(\tau>t_2)=1$
\bea \label{ex12} 
- \partial_\tau \log \tilde g(\tau) = b \tilde g(\tau) + \frac{a}{\tilde g(\tau)} - (b+a) + \ln h_2 \delta(\tau-t_2)
+ \ln f_1 \delta(\tau-t_1)
\eea
One gets $\tilde g(t_2^-)= h_2$, then one solves backwards from $\tau=t_2^-$ to $\tau=t_1^+$,
which amounts to solve forward for $g(t)=\tilde g(t_2-\tau)$
\bea
 \partial_t \log g(t) = b g(t) + \frac{a}{g(t)} - (b+a)  \quad , \quad g(0)=h_2
\eea
which is equivalent to \eqref{KPPsimple}, and one gets 
$\tilde g(t_1^+) = g_{h_2}(t_2-t_1)$. Going back to \eqref{ex12} this leads to
$\tilde g(t_1^-) = h_1 g_{h_2}(t_2-t_1)$. Finally solving down to $\tau=0$ we
obtain $\tilde g(0)=g_{h_1 g_{h_2}(t_2-t_1)}(t_1)$ which is the desired result. 
\\

We can generalize correlation functions of the type \eqref{prod1} to an observable which
probes the process $M(t)$ all along its trajectory. We then define
\be
g(t) = \mathbb{E}_1(e^{ \int_0^t M(t') \log h(t') })  \quad , \quad \tilde g(\tau)=g(t-\tau) 
\ee
and taking a limit of the delta function sources in \eqref{sourceshj} it is easy to see that
$\tilde g(\tau)$ now satisfies the backward equation
\bea
- \partial_\tau \log \tilde g(\tau) = b \tilde g(\tau) + \frac{a}{\tilde g(\tau)} - (b+a) + \log h(\tau) 
\eea
Again, it should be solved backward in time starting from $\tilde g(\tau>t)=0$ down to
to $\tau=0$, and then one retrieve $g(t) = \tilde g(0)$. 
\\

{\bf Remark}. Equivalently one can state that to calculate
\be
g(t_n) = \mathbb{E}_1(e^{ \int_0^{t_n} M(t') \log h(t') }) = \mathbb{E}_1(e^{ \int_0^{t_n} M(t_n-t') \log h(t_n - t') }) 
\ee
one solves forward 
\bea
\partial_t \log g(t) = b g(t) + \frac{a}{g(t)} - (b+a) + \log h(t_n - t) 
\eea
Similarly for the discrete observables \eqref{prod1} one can instead solving forward
\bea
 \partial_t \log g(t) = b g(t) + \frac{a}{g(t)} - (b+a) + \sum_j \ln f_j \delta(t-(t_n-t_j))
\eea


\subsection{Multi-time observables, general case} 

It is simple to extend the previous calculation to include space. Consider the BBM where each particle
undergoes Brownian motion with $dx_i^2= 2 D dt$. Let us start from the standard 
single time generating function
\be
g(x,t)= g_{h(\cdot)}(x,t) = \mathbb{E}_{1,\{x \} } \left( \prod_{j=1}^{M(t)} h(x_j(t)) \right)  \label{defg2} 
\ee 
which is the solution for $t \geq 0$ of 
\be  \label{kpp12} 
 \partial_t g = D \nabla_x^2 g + b g^2 + a g - (b+a) g  \quad, \quad g(x,t=0)=h(x) 
\ee
We recall that $\mathbb{E}_{1,\{x \} }$ means expectation value given that the process starts at $t=0$ with a single particle 
$M(0)=1$ at $x_1(0)=x$. 

Then one considers the the following two time 
generating function, which again can be expressed from the single time one, $g_{h(\cdot)}(x,t)$, as follows
\bea \label{2timespace} 
&& g(x,t_1,t_2)= \mathbb{E}_{1,\{x \} }  \left( \prod_{j=1}^{M(t_1)} h_1(x_j(t_1)) 
\prod_{j=1}^{M(t_2)} h_2(x_j(t_2))  \right)   \\
&& = \mathbb{E}_{1,\{x \} }  \left( \prod_{j=1}^{M(t_1)} h_1(x_j(t_1)) 
\prod_{j=1}^{M(t_1)} g_{h_2(\cdot)}(x_j(t_1),t_2-t_1) \right) =
g_{h_1(\cdot) g_{h_2(\cdot)}(t_2-t_1)}(x,t_1)
\eea

To obtain information over the whole process in time we thus define a more general 
generating function depending on the test function $h(x,t)$ 
\be \label{gg} 
g(x,t) = \mathbb{E}_{1,\{x \} } (e^{ \int_0^t dt' \sum_{j=1}^{M(t')} \log h(x_j(t'),t') }) 
\ee
For the choice $\log h(x,t)=\sum_{j=1}^n \delta(t-t_j) \log h_j(x)$ and $n=2$ it reduces to \eqref{2timespace}. 
Define $\tilde g(x,\tau)=g(x,t-\tau)$,
then $\tilde g(x,\tau)$ satisfies
\bea
- \partial_\tau \log \tilde g(x,\tau) = \frac{D}{ \tilde g(x,\tau)} \nabla_x^2 \tilde g(x,\tau) + b \tilde g(x,\tau) + \frac{a}{\tilde g(x,\tau)} - (b+a) + \log h(x,\tau) \label{eqh} 
\eea
which one solves backward starting from $\tilde g(x,\tau>t)=1$ down to
to $\tau=0$, and then one retrieve $g(x,t) = \tilde g(x,0)$. 
\\

Denoting $\Lambda(x,t)=\log h(x,t)$, we can formulate the following statement 
for expectation values of the BBM, in the form given in the text and for a general
branching mechanism. The
generating function
\be
g(x) = \mathbb{E}_{1,\{x \} } (e^{ \int_0^{+\infty} dt' \sum_{j=1}^{M(t')} \Lambda(x_j(t'),t') })  
\ee
is obtained by solving for $\tilde g(x,\tau)$ 
\bea \label{eqLambda} 
- \partial_\tau \tilde g(x,\tau) = D \nabla_x^2 \tilde g(x,\tau) + V ( \Phi(\tilde g(x,\tau)) - 
\tilde g(x,\tau)) +  \Lambda(x,\tau) \tilde g(x,\tau) 
\eea
backward in time starting from $\tilde g(x,\tau=+\infty)=1$ down to
to $\tau=0$, and then one retrieve $g(x) = \tilde g(x,0)$. We have pushed 
the uppper bound $t$ to infinity in \eqref{gg}, so we assume that 
$\Lambda(x,t)$ vanishes beyond some time. 
This formula can be generalized to an arbitrary initial condition
with $M(0)$ particles at positions $x_j(0)$ as
\be \label{genfin} 
\mathbb{E}_{M(0),\{x_j(0)\}}(e^{ \int_0^{+\infty} dt \sum_{j=1}^{M(t)} \Lambda(x_j(t),t) })  
= e^{ \sum_{j=1}^{M(0)} \ln g(x_j(0)) }
\ee
\\

{\bf Limit to the BFM instanton equation}. We can go then directly from \eqref{eqh}, \eqref{eqLambda} 
to the BFM instanton equation by  introducing a parameter $N$, which will be taken large, and
choosing $M(0)=O(N)$ as well as
\bea
&&  \log h(x,\tau) = \Lambda(x,\tau)  = \frac{1}{N} \lambda(x,\tau) 
\quad , \quad  \tilde g(x,\tau) = 1 + \frac{1}{N} \tilde u(x,\tau) + O(\frac{1}{N^2}) 
\eea 
One sets $D=1$. For the binary model one scales $b=\tilde b N$ and $d=\tilde b N - \gamma$ with $\gamma=-m^2$,
$\tilde b=\sigma$. More generally one scales $V \simeq 2 \tilde b N = 2 \sigma N$,  
$\Phi'(1)-1=\frac{\gamma}{V}=- m^2/V$ and $\frac{1}{2} \frac{V}{N} \tilde \Phi''(1)=\tilde b=\sigma$.
Then one obtains in the large $N$ limit the instanton equation of the BFM
\bea
- \partial_\tau \tilde u(x,\tau) = \nabla_x^2 \tilde u(x,\tau) + \sigma \tilde u(x,\tau)^2 
- m^2 \tilde u(x,\tau) + \lambda(x,\tau) 
\eea
which is also solved backwards in time. With the identification $\frac{1}{N} \sum_{j=1}^{M(t)} \delta(x-x_j(t) \to \dot u(x,t)$ 
the equation \eqref{genfin} becomes in the large $N$ limit 
\be 
\mathbb{E}_{\dot u(\cdot,0)}(e^{ \int_0^{+\infty} dt \int d^d x \lambda(x,t) \dot u(x,t) })  
= e^{ \int d^d x \dot u(x,0) \tilde u(x,0) } 
\ee 
This is a particular case of the identity \eqref{generating} for the BFM, corresponding
to the kick at $t=0$, $\dot f(x,t)= \dot u(x,0) \delta(t)$ (with $\eta=1$). This kick creates
the appropriate density of BBM particles at $t=0$. 
\\

{\bf BFM with immigration}. 
By analogy with the BFM, one can go beyong the kick and describe also the BBM in presence of immigration.
Denoting $N \rho(x,t)= \sum_{j=1}^{M(t)} \delta(x-x_j(t))$, one assumes that in each time slice $dt$ 
there is an addition of an atomic measure described by $N d \rho(x,t)=\dot f(x,t) dt$ 
(it is atomic since in the BBM we are adding particles at some locations). One then has
\be
\mathbb{E}(e^{ \int_0^{+\infty} dt \sum_{j=1}^{M(t)} \Lambda(x_j(t),t) })  
= e^{\int_0^{+\infty} dt d^dx \dot f(x,t) \ln \tilde g(x,t) }
\ee
which generalizes \eqref{genfin} (note that here $N$ is arbitrary and can be taken to be $N=1$). 
 
%
%
%
%
%

%
%
%

\section{Immigration driven by a Levy process}
\label{app:immigration} 

Here we give a simple result for avalanches in $d=0$ driven by a Levy process.
It applies equivalently to the ABBM model driven by a time dependent force $f(t)$
which is a Levy process in time, or as a Feller process with immigration rate $\dot f(t)$. 

Let us recall that a L\'evy process \cite{bertoinLevy,bertoin2}
is a real random function $X(w)$, continuous on the right with a limit on the left, i.e. it can have jumps. It has homogeneous and independent increments, i.e. $\{X (w_{i+1} ) - X (w_i ) \}_{i=1,...,p}$ are independent random variables for any $w_1  < ... < w_p$ and any $p$, and for all $w < w'$
the law of  $X(w')-X(w)$ is the same as the law of $X(w' - w) - X(0$). Its characteristic function satisfies, for $w > 0$
\bea \label{Levyexp} 
 \langle e^{\omega (X(w)-X(0))} \rangle = e^{w \phi(\omega)} \quad , \quad  \langle e^{- \int dw \omega'(w) X(w)} \rangle = e^{\int dw \phi(\omega(w))} 
\eea
where $\phi(\omega)$ is the L\'evy exponent, with $\phi(0)=0$. Here $\omega(w)$ is a function
which vanishes at infinity \cite{bertoin2}. In general the L\'evy exponent can be written as
\bea
\phi(\omega) = a \omega + \frac{b}{2} \omega^2 + \int_{s} ds n(s) (e^{s \omega}-1- s \omega \theta(|s|<1)) 
\eea
where $n(s) ds$ is the jump measure with $\int ds n(s) \min(1,s^2) < +\infty$. 

Consider now the ABBM model with a driving force $f(t)$. One can study 
averages over the process $f(t)$ by averaging over both sides in the
definition of the generating function \eqref{generating} (in $d=0$ with $f(x,t)=f(t)$). One has (with $\dot u(0)=0$)
\be
 \langle G[\lambda] \rangle_f := \langle \overline{e^{\int_0^{+\infty} dt  \lambda(t) \dot u(t)}} \rangle_f
= \langle e^{ \int_0^{+\infty} dt  \dot f(t) \tilde{u}^\lambda(t) } \rangle_f  
= \langle e^{ - \int_0^{+\infty} dt  f(t) \partial_t \tilde{u}^\lambda(t) } \rangle_f
\ee
assuming that $[f(t) \tilde{u}^\lambda(t)]_0^{+\infty}=0$. 
Let us now choose $f(t)$ to be a Levy process ($f = X$ and $t=w$)
with only positive jumps (since we want monotonous driving). For simplicity we write its Levy exponent
as 
\be 
\langle e^{\omega (f(t)-f(0)} \rangle = e^{t \phi(\omega)} \quad , \quad 
\phi(\omega) = v \omega + \int_{s>0} ds \, n(s) (e^{s \omega} -1- s \omega)  
\ee
assuming that $n(s)$ is such that all integrals converge. Here $v$ is the driving velocity. Using \eqref{Levyexp} we find
\bea \label{106} 
&&  \langle G[\lambda] \rangle_f = 
e^{  \int_0^{+\infty} dt  \phi(\tilde{u}^\lambda(t)) } = 
e^{  \int_0^{+\infty} dt \left(  v \tilde{u}^\lambda(t) +  \int_{s>0} ds n(s) (e^{s \tilde{u}^\lambda(t)}-1-s \tilde{u}^\lambda(t))  \right) } 
\eea
The simplest application is to compute $\langle \overline{ e^{\lambda \dot u(t_0) } } \rangle_f$ 
which amounts to insert the solution of the instanton equation \eqref{instanton} in $d=0$,
with a source $\lambda(t)=\lambda \delta(t-t_0)$,
i.e. $\tilde u^\lambda(t) = \frac{\lambda}{\lambda + (1- \lambda) e^{-(t-t_0)}}$. In the absence of 
jumps, i.e. $f(t)=v$, $n(s)=0$, one recovers for $t_0 \to +\infty$ and upon Laplace inversion, the stationary PDF 
$P(\dot u)= (\dot u)^{-1+v} e^{- \dot u}/\Gamma(v)$ (for $m=\sigma=\eta=1$)
see e.g. \cite[SecIIIA]{DobrinevskiLeDoussalWiese2011b} or \cite{LeDoussalWiese2011a}. The Eq. \eqref{106}
allows to compute the deviations from this PDF due to the jumps. We leave its study to the future. 

There are recent related calculations in probability theory, see \cite{HeImmigration2016}.
For classical works on CB processes with immigration see \cite{WatanabeImmigration,KellerImmigration2012}, 
and \cite{Pakes1972} for the BGW model with immigration.

\section{Duality}
\label{app:duality} 

Two Markov processes $X(t)$ and $Y(t)$ are dual w.r.t. a function $H(X,Y)$ if (see e.g. 
\cite[Thm1.23]{Etheridge}) 
\be
\mathbb{E}_{X(0)}[ H(X(t),Y(0)) ] = \mathbb{E}_{Y(0)}[ H(X(0),Y(t)) ] 
\ee
or, in terms of generators $L_X H(\cdot , y)(x) = L_Y H(x,\cdot)(y)$. 
One classical example \cite{DualityReview,DualityPolynomialsGiardina} 
is the Wright-Fisher diffusion  
\be
dX_t = \sqrt{X_t(1-X_t)} dB_t \quad , \quad L_{WF} = \frac{1}{2} x(1-x) \partial_x^2 
\ee 
which is dual to the (integer) block-counting process $N_t$ of Kingman's
coalescent \cite{Kingman} (where 2 blocks are merged at random) 
of generator 
\bea
L _K f(n) = \frac{n(n-1)}{2} (f(n-1) - f(n)) 
\eea 
The (moment) duality function is $H(x,n)=x^n$. Indeed one checks that
\be 
L_K x^n = L_{WF} x^n =  \frac{n(n-1)}{2} (x^{n-1}-x^n) 
\ee 
This implies the duality relation 
\bea
\mathbb{E}_x[ X_t^n ] = \mathbb{E}_n[ x^{N_t} ] 
\eea 
It gives interesting information, for instance for $n=1$, 
$\mathbb{E}_x[ X_t] = 1- {\rm Prob}(X_\infty=0|X_0=x) = \mathbb{E}_1[x]=x$.

The SBM also satisfies a duality relation, indeed the property \eqref{duality1}
says that the SBM is dual to a deterministic process, whose time evolution is described
by the "instanton equation" \eqref{instantonSBM}, see \cite[Sec 1.6]{Etheridge}.

\section{Superprocesses}
\label{app:superprocesses} 

Jirina, Watanabe and others developped the theory of continuous state branching processes (CB-processes),
see \cite{Jirina,Skorohod64,Watanabe68,LampertiCB,lamperti,MotooCB}. A measured value CB process \cite{Watanabe68}
$\rho_t$ satisfies the so-called branching property
\bea
\mathbb{E}_{\mu_1+\mu_2}\left[ e^{- (\rho_t , \phi)} \right] = \mathbb{E}_{\mu_1} \left[e^{- (\rho_t , \phi)} \right]
\mathbb{E}_{\mu_2}\left[ e^{- (\rho_t , \phi)} \right]
\eea 
for any initial measures $\mu_1,\mu_2$ and "positive" function $\phi$. It means
that if $\rho^{(1)}_t$ and $\rho^{(2)}_t$ are SBM then $\rho^{(1)}_t+\rho^{(2)}_t$ is also a SBM
with initial condition
sum of the two. In \cite{Watanabe68} it is shown that the duality relation holds
\bea
\mathbb{E}_{\rho_0}\left[ e^{- (\rho_t,\phi) } \right] = e^{- \int \psi_t(x;\phi) \rho_0(dx) } 
\eea 
and that there is a one to one correspondence between (regular) 
CB processes and $\Psi$-semigroup $\psi_t$, i.e. which
obey $\psi_{t+s}(x,\phi)=\psi_t(x,(\psi_s(.,\phi))$, $\psi_0(x,\phi)=\phi(x)$.
These are related to infinitely divisible measures $P$ through 
$L_P(\phi)=e^{-\psi_{t}(x,\phi)}$ where $L_P(\phi) = \int P(dx) e^{- \phi(x) }$. 
This leads to a very general "instanton equation"
\bea
\partial_t \psi_t = A \psi_t + \sigma( \phi(\cdot;\psi_t) - \psi_t) \quad , \quad \psi_0 = f 
\eea 
where $A$ is a linear operator, $\sigma(x)$ a non-negative function, and $\phi(x;f)$ a $\Psi$-function. 
Semi groups exist in the cases $\partial_t \psi_t = A \psi_t  - \sigma \psi_t^\alpha$, $1<\alpha \leq 2$ and
$\partial_t \psi_t = A \psi_t  + \sigma \psi_t^\alpha$ for $0<\alpha \leq 1$, $\sigma>0$. 
Some examples were discussed in the main text. 

As mentionned in the text there is a "Levy-Kintchine like" representation for
the characteristic function of an infinitely divisible random measure $\mu$
in terms of $\mu_d$ and $m(d\nu)$ (see details in \cite[Thm1.28]{Etheridge}) 
\be
- \ln \mathbb{E}( e^{- \langle \mu,\phi \rangle}) = \langle \mu_d,\phi \rangle + \int m(d \nu) (1- 
e^{- \langle \nu,\phi \rangle}) 
\ee
so that the duality \eqref{duality1} holds while the SBM "instanton equation" \eqref{instantonSBM}
for $v(x,t)$ 
is generalized into 
\be 
\partial_t v = (A  v)(x,t)  - \tilde b(x) v^2 + \gamma(x) v + \int_0^{+\infty} n(x,d\theta) (1- e^{- \theta v} - \theta v)   
\ee 
where we also allowed for spatial inhomegeneity. Here $n$ is a positive "Levy jump" kernel
(see details in \cite[Sec1.7]{Etheridge}).


Finally note that there is a one-to-one correspondence between continuous-state branching processes 
and L\'evy processes with no negative jumps under a time change \cite{LampertiCB}, see e.g. \cite{Simatos,Caballero,PatieLevyCBImmigration}.


\end{appendix}

\bibliographystyle{iopart-num}

\begin{thebibliography}{100}
\bibitem{footnote2}
provided the initial condition is such that $\dot u(x,t=0) \geq 0$.
\bibitem{footnoteC} Eq. \eqref{instanton} should be solved backward in time, with
$\tilde u(x,+\infty)=0$. For a source $\lambda(x,t)$ vanishing at large $|x|$, $\tilde u(x,t)$
should also vanish.
\bibitem{footnoteN} The SBM measures, including $\rho_0$, are usually not normalized
to a total mass equal to unity, so here $N$ is an independent parameter, 
in general distinct from $M(0)$.
\bibitem{footnoteHit} There are subtelties defining "meets a ball". "Ever charges set $A$" means
that $\rho_t(A)>0$ for some $t$, while "hits $A$" means ${\cal R} \cap A \neq 0$,
see more detailed results in \cite[Sec. 6.5]{Etheridge}.
\end{thebibliography}

\end{document}